\documentclass[twocolumn,amsmath,showpacs,amstex,eqsecnum,pre,aps]{revtex4}

\usepackage{amsmath}
\usepackage{graphicx}
\usepackage{dcolumn}
\usepackage{bm}
\usepackage{float}
\usepackage{subfigure}
\usepackage[usenames]{color}
\usepackage{amssymb}
\usepackage{txfonts}
\usepackage{wasysym}

\usepackage[
    bookmarks=true,
    bookmarksopen=true,
    colorlinks=true,
    anchorcolor=blue,
    filecolor=green,
    citecolor=blue,
    urlcolor=blue,
    linkcolor=red,
    menucolor=blue,
    breaklinks=true
]{hyperref}

\begin{document}
\draft

\title{Self-consistent generalized Langevin equation theory of the
dynamics of multicomponent atomic liquids}
\author{Edilio L\'azaro-L\'azaro$^{1}$, Patricia Mendoza-M\'endez$^{1}$,
Luis Fernando Elizondo-Aguilera$^{2,3}$, Jorge Adri\'an Perera-Burgos$^4$, 
Pedro Ezequiel Ram\'irez Gonz\'alez$^{5}$, Gabriel P\'erez-\'Angel$^6$,
Ram\'on Casta\~neda-Priego$^{2}$ and Magdaleno Medina-Noyola$^{1,2}$}
\affiliation{$^{1}$ Instituto de F\'{\i}sica {\sl ``Manuel Sandoval Vallarta"},
Universidad Aut\'{o}noma de San Luis Potos\'{\i}, \'{A}lvaro
Obreg\'{o}n 64, 78000 San Luis Potos\'{\i}, SLP, Mexico}
\affiliation{$^{2}$ Departamento de Ingenier\'ia F\'isica, Divisi\'on de Ciencias
e Ingenier\'ias, Universidad de Guanajuato, Loma del Bosque 103, 37150 Le\'on, Mexico.}
\affiliation{$^{3}$ Institut f\"ur Materialphysik im Weltraum, Deutsches Zentrum
f\"ur Luft-und
Raumfahrt (DLR), 51170 K\"oln, Germany}
\affiliation{$^{4}$ Facultad de Ciencias Qu\'imica y Petrolera, Universidad
Aut\'onoma del Carmen, C. 56 No.4 Esq. Avenida Concordia, Col. Benito Ju\'arez,
C.P. 24180, Cd. del Carmen, Campeche, Mexico.}
\affiliation{$^{5}$ CONACYT Research Fellow-\ Instituto de F\'{\i}sica {\sl ``Manuel
Sandoval Vallarta"}, Universidad Aut\'{o}noma de San Luis Potos\'{\i}, \'{A}lvaro
Obreg\'{o}n 64, 78000 San Luis Potos\'{\i}, SLP, M\'{e}xico}
\affiliation{$^{6}$ Departamento de F\'isica Aplicada, Cinvestav, Unidad M\'erida,
Apartado Postal 73 Cordemex,
97310 M\'erida, Yucat\'an, Mexico }

\date{\today}

\begin{abstract}
A fundamental challenge of the theory of liquids is to understand the similarities
and differences in the macroscopic dynamics of both colloidal and atomic liquids, 
which originate in the (Newtonian or Brownian) nature of the microscopic motion of
their constituents. Starting from the recently-discovered long-time dynamic equivalence
between a colloidal and an atomic liquid that share the same interparticle pair potential, 
in this work we develop a self-consistent generalized Langevin equation (SCGLE) theory for 
the dynamics of equilibrium multicomponent atomic liquids, applicable as an approximate but
quantitative theory describing the long-time diffusive dynamical properties of simple 
equilibrium atomic liquids. When complemented with a Gaussian-like approximation, this 
theory is also able to provide a reasonable representation of the passage from ballistic to 
diffusive behavior. We illustrate the applicability of the resulting theory with three particular 
examples, namely,  a monodisperse and a polydisperse monocomponent  hard-sphere liquid, and a highly
size-asymmetric binary  hard-sphere mixture. To assess the quantitative accuracy of our results, we
perform event-driven molecular dynamics simulations, which corroborate the general features of the 
theoretical predictions.

\end{abstract}

\pacs{23.23.+x, 56.65.Dy}

\maketitle

\section{Introduction}\label{sectionI}
It is well known that the structural and dynamical
properties of atomic liquids and colloidal fluids exhibit an
almost perfect correspondence \cite{pusey0,deschepperpusey1,deschepperpusey2}.
This analogy seems to be particularly accurate regarding the rather
complex dynamical behavior of both systems as they approach the glass
transition \cite{lowenhansenroux,szamelflenner,puertasaging,szamellowen}.
While the similarity of the equilibrium phase behavior of colloidal and
atomic systems with analogous interactions is well understood,
determining the range of validity of such correspondence at the level
of dynamical properties remains a relevant topic in the
study of the dynamics of liquids.

In a recent contribution \cite{atomic0}, the generalized Langevin equation
(GLE) formalism \cite{faraday,delrio} was employed to derive the exact equation of
motion of individual tracer particles \cite{atomic1}, as well as the exact
time-evolution equations for both collective and self intermediate scattering
functions, $F(k, t)$ and $F_S(k,t)$, respectively, of an atomic liquid \cite{atomic2},
with $k$ being the magnitude of the wavevector. A remarkable fundamental
consequence of these theoretical results is the general prediction
that, properly scaled, the strictly \emph{long-time} dynamics of an atomic liquid
should be indistinguishable from the dynamics of the Brownian liquid with the same
interparticle interactions. This prediction has been successfully tested by
computer simulations \cite{atomic1,atomic2,atomic3}.

The first main purpose of the present work is to adapt now the arguments and
approximations previously employed in the proposal of the approximate SCGLE 
theory of \emph{colloid} dynamics \cite{scgle1,scgle2} and dynamical arrest 
\cite{todos1,todos2}, to convert the exact results for tracer diffusion in 
Ref.  \cite{atomic1} and for  $F(k, t)$ and $F_S(k,t)$ in Ref. \cite{atomic2},
into an approximate theory for the \emph{long-time} dynamic properties of a 
monocomponent atomic liquid.  The  expectation is that the resulting atomic 
extension of the SCGLE theory will provide an unifying theoretical framework
to describe in more detail the most relevant similarities and differences 
between the macroscopic dynamics of atomic and colloidal liquids (in the 
absence of hydrodynamic interactions). This implies that in general, if the 
dynamical properties of a Brownian fluid whose molecules interact with a given 
interaction potential have been explicitly determined, then one has automatically
determined the long-time dynamics of its equivalent atomic system. This, however,
involves the proper determination \cite{atomic0} of the  ``short-time" self-diffusion
coefficient $D^0$ of the atomic liquid by simple random-flight and kinetic-theoretical
arguments \cite{mcquarrie}. 

Although irrelevant for the study of phenomenology such as 
the slow dynamics of glass-forming liquids, 
a conceptually important issue is the difference in the short-time dynamics of 
colloidal and atomic liquids: the dynamics of the former is diffusive at all relevant 
timescales, whereas the dynamics of the latter crosses over from ballistic to diffusive 
after a few particle collisions. Thus, a secondary aim of this paper is to show how this 
SCGLE theory for the long-time dynamics of atomic liquids, may be complemented with a
short-time Gaussian approximation in order to provide a reasonable and simple representation 
of the passage from ballistic to diffusive behavior of the mean-square displacement and the 
intermediate scattering functions of the atomic liquid. With the aim of testing the main 
features and the quantitative accuracy of the resulting  ``first principles'' approximate
theory of the dynamics of atomic liquids, in this paper we also present the results of a 
set of event-driven molecular dynamics simulations for the hard-sphere model liquid.

The second main purpose of this paper is to use the multicomponent extension of the 
SCGLE theory of colloid dynamics \cite{marco2,rigo1} to further extend the SCGLE theory
of the dynamics of monoatomic liquids just described, now to \emph{mixtures} of \emph{atomic}
liquids. This opens the possibility to model the dynamics of a large class of scientifically 
and technologically relevant systems and materials, many of which present interesting 
glass-forming properties. In this regard let us emphasize that the present multicomponent SCGLE 
theory shares these aims with the well-known mode coupling theory (MCT) of the ideal glass
transition \cite{goetze1,goetze2} and its multi-component extension \cite{bossethakur1, barratlatz1,NAGELE2}. 

Unfortunately, MCT and the SCGLE theory are strictly theories of the dynamic properties of 
liquids in thermodynamic equilibrium. Thus, they are unable to predict the most interesting 
and essential non-equilibrium features of glassy states, such as aging, and the dependence 
of the properties of glassy materials on their preparation protocol \cite{angellreview1}. The 
SCGLE theory, however, was recently extended to genuine non-equilibrium conditions \cite{nescgle1},
thus contributing to demolish this severe limitation. In its first applications, the resulting
\emph{non-equilibrium} self-consistent generalized Langevin equation (NE-SCGLE) theory of
irreversible processes in liquids has exhibited a remarkable predictive power, particularly in
setting the description of the kinetics of the aging of glasses and gels in the same conceptual 
framework using simple Lennard-Jones--like benchmark models \cite{nescgle3,nescgle5}.
Applying this new non-equilibrium theory to multicomponent atomic liquids, however, has as a
prerequisite the previous development of the equilibrium version of the SCGLE theory, and this 
provides another fundamental reason for our present study, which will thus be strictly confined to
equilibrium conditions.

In order to test the reliability of our proposed SCGLE theory, here we also compare its predictions
with the results of our event-driven molecular dynamics simulation involving two illustrative 
applications: a polydisperse HS liquid, modeled as a moderately size-asymmetric binary HS mixture,
and a genuine, highly size-asymmetric, binary HS mixture. As the corresponding comparisons indicate,
the present equilibrium SCGLE theory of the dynamics of multicomponent atomic liquids does provide the 
correct qualitative features and a very acceptable quantitative description of the dynamics of these systems.

This manuscript is organized as follows. In section \ref{sectionII}, the
main results of the derivation of three general time-evolution equations are
summarized. In the same section, we also introduce those approximations that
transform these equations into a fully self-consistent system of equations for
the overdamped dynamics of a monocomponent atomic liquid. As mentioned above,
such an overdamped SCGLE equations do not describe properly the ballistic behavior
of the atomic liquid. Thus, in Sec. \ref{sectionIII}, a simple approximation to
fully account for the ballistic regime is explored within the so-called Gaussian
approximation; this approach will allow us to describe the passage time of
$F_{S}(k,t)$ and the mean squared displacement (MSD) from the ballistic to the diffusive time regime. Although
the results described in sections \ref{sectionII} and \ref{sectionIII} are only
applicable to monocomponent atomic liquids, the multicomponent extension of the
SCGLE theory for the dynamics of atomic liquids is presented in section \ref{sectionIV}.
The predictions of the resulting SCGLE theory for both, monocomponent and
multicomponent atomic liquids, are discussed and compared in detail with event-driven
computer simulations in sections \ref{sectionIII} and \ref{sectionIV}, respectively.
The main conclusions of this work are finally summarized in Sec. \ref{sectionV}.

\section{Review of the tracer, collective, and self-diffusion in \emph{monocomponent}
atomic liquids.}\label{sectionII}

In this section we briefly review the main concepts and results
of Refs. \cite{atomic1} and \cite{atomic2}, upon which we build the
approximate SCGLE theory of the \emph{long-time} dynamics of
\emph{monocomponent} atomic liquids. As we mentioned in the introduction,
in Ref. \cite{atomic1} the generalized Langevin equation formalism
\cite{delrio,faraday} was applied to derive the stochastic time-evolution
equation for the velocity $\textbf{v}(t)$ that describes the Brownian motion
of individual tracer particles in an atomic liquid. In Ref. \cite{atomic2},
general memory-function equations were derived for the intermediate
scattering functions (ISFs) $F(k,t)$ and $F_S(k,t)$, which describe
the collective and self motion, respectively, of atomic liquids. The overdamped version of these exact
equations are formally identical to the corresponding equations of
a Brownian liquid. Therefore, in this section, we introduce the same
approximations employed before in the construction of the approximate
SCGLE theory of Brownian liquids \cite{scgle1,scgle2,todos1,todos2} in order to
build the SCGLE theory for the long-time dynamics of monocomponent atomic liquids.

\subsection{Brownian motion of atomic tracers} \label{subsectionII.1}
Let us consider a simple atomic fluid formed by $N$ identical spherical
particles in a volume $V$ at a temperature $T$, whose microscopic dynamics
is described by Newton's equations,

\begin{equation}
M{\frac{d{\bf v}_{\alpha}(t)}{dt}}= \sum_{\beta\neq \alpha}{\bf F}_{\alpha\beta}(t),\quad
(\alpha=1,2,\ldots ,N), \label{eq1}
\end{equation}
where $M$ is the mass and ${\bf v}_{\alpha}(t)=d{\bf r}_{\alpha}(t)/dt$ the
velocity of the $\alpha$th particle at position ${\bf r}_{\alpha}(t)$, and in
which the interactions between the particles are represented by the
sum of the pairwise forces, with ${\bf F}_{\alpha\beta}=-\nabla_\alpha
u(|{\bf r}_{\alpha}-{\bf r}_{\beta}|)$ being the force exerted on particle
$\alpha$ by particle $\beta$ and $u(|{\bf r}_{\alpha}-{\bf r}_{\beta}|)$ the pair
potential among particles. In Ref. \cite{atomic1}, the general aim
was to establish a connection between the microscopic dynamics
described by these (Newton's) equations and the macroscopic
dynamical properties of the atomic fluid.

The main result of Ref. \cite{atomic1} is the derivation of the
generalized Langevin equation that describes the ballistic-to-diffusive
crossover of a tagged particle in the atomic liquid. Such stochastic
equation reads

\begin{equation}
M{\frac{d{\bf v}(t)}{dt}}= -\zeta^0 {\bf v}(t)+{\bf f}
^0(t)- \int_0^t dt' {\Delta \zeta(t}-t') {\bf v}(t')+
{\bf F} (t). \label{eq6p}
\end{equation}
In this equation, the random force ${\bf f}^0 (t)$ is a
``white" noise, i.e., a  delta-correlated, stationary and Gaussian
stochastic process with zero mean and time-dependent
correlation function given by $\overline{{\bf f}^0(t){\bf f}^0(t')}
=\stackrel{\leftrightarrow }{{\bf I}}k_BT\zeta^0 2\delta(t-t')$,
with $\stackrel{\leftrightarrow }{{\bf I}}$ being the 3$\times$3
cartesian unit tensor and $k_BT$ the thermal energy. The term involving
the time-dependent friction function $\Delta\zeta(t)$ describes the average
dissipative friction effects due to the conservative (or ``direct")
forces on the tracer particle, whose random component is the stationary
stochastic force $\textbf{F}(t)$ that obeys the fluctuation-dissipation relationship  $\overline{{\bf F}(t)
{\bf F}(t')}=\stackrel{\leftrightarrow }{{\bf I}}k_BT \Delta \zeta (t-t')$.
Thus, the configurational effects of the interparticle interactions
are embodied in the time-dependent friction function $\Delta \zeta (t) $,
for which the following approximate expression was also derived in Ref. \cite{atomic1},
\begin{equation}
\Delta \zeta (t) =\frac{k_BT}{3\left( 2\pi \right) ^{3}n}\int d
{\bf k}\left[\frac{ k[S(k)-1]}{S(k)}\right] ^{2}F(k,t)F_S(k,t),
\label{dzdt0p}
\end{equation}
where $n\equiv N/V$ is the number density and $S(k)$ is
the static structure factor.

Eqs. (\ref{eq6p}) and (\ref{dzdt0p}), valid for an atomic liquid,
turn out to be formally identical to the corresponding results derived
in Ref. \cite{faraday} for a Brownian liquid in the absence of
hydrodynamic interactions, see, e.g.,  Eqs. (4.9) and (4.10) of Ref.
\cite{faraday}. The most remarkable conclusion of Ref.
\cite{atomic1} is that the Brownian motion of any labeled particle
in an atomic liquid is formally described by the same equation that
describes the Brownian motion of a labeled particle in a liquid of
interacting colloidal particles, provided both liquids share the same
thermodynamic conditions and their molecular constituents interact
with the same kind of pair potential. The fundamental difference between
these two dynamically distinct systems lies in the physical origin
of the friction force  $-\zeta^0{\bf v}(t)$ and in the determination
of the friction coefficient $\zeta^0$: in a colloidal liquid the
friction force $-\zeta ^{0}{\bf v}(t)$ is caused by the supporting
solvent, and hence, $\zeta^{0}$ assumes its Stokes value \cite{mcquarrie}.
In contrast, the friction force  $-\zeta^0{\bf v}(t)$ in an atomic liquid is not
caused by any external material agent, instead, its origin is a more subtle kinetic
mechanism \cite{atomic0}, which originates in the spontaneous
tendency to maintain the equipartition of kinetic energy through
molecular collisions. Uhlenbeck and Ornstein refer to this effect
as \emph{``Doppler"} friction,  caused by the fact that when any
tracer particle of the fluid ``\emph{is moving, say to the right, will
be hit by more molecules from the right than from the
left}" \cite{ornsteinuhlenbeck}. These collisional effects are described
by the kinetic friction coefficient
$\zeta^0$ which, according to Refs. \cite{atomic0,atomic1}, is defined
through Einstein's relation,
\begin{equation}
\zeta^0\equiv k_BT/D^0,
\label{einstein}
\end{equation}
with the short-time self-diffusion coefficient $D^0$ determined
by the same arguments employed in the elementary kinetic theory of gases
\cite{mcquarrie}. According to such arguments, the mean free path $l_0$
can be approximated as $l_0\approx 1/n\sigma^2_{col}$, with $\sigma_{col}$
being the collision diameter of the atoms. The diffusion coefficient that
results from a large number of successive collisions is thus given by
$D^0=l_0^2/\tau_0$, where $\tau_0$ is the mean free time, related to
$l_0$ by $l_0/\tau_0=v_0\equiv \sqrt{k_BT/M}$. The resulting
value of $D^0$ is given by \cite{mcquarrie}
\begin{equation}
D^0\equiv \frac{3}{8
}\left(\frac{k_BT}{\pi M}\right)^{1/2}\frac{1}{n\sigma^2_{col}}.
\label{dkinetictheory}
\end{equation}

\subsection{Collective and self-diffusion in atomic liquids}\label{subsubsectionII.2}

In  Ref. \cite{atomic2}, the GLE formalism \cite{faraday,delrio}
was also employed to derive the exact time-evolution equations for
the collective and self intermediate scattering functions,
$F(k, t)$ and $F_S(k,t)$, of an atomic liquid in terms of the
corresponding memory functions, see, e.g., Eqs. (32) and (33) of
Ref. \cite{atomic2}. For times $t$ sufficiently long compared with
$\tau_0$, i.e., the so-called ``overdamped'' limit, and in terms of
the corresponding Laplace transforms $F(k,z)$ and $F_S(k,z)$, these
equations read (see, e.g., Eqs. (37) and (38) of Ref. \cite{atomic2})

\begin{equation}
F(k,z) = \frac{S(k)}{z+\frac{k^{2}S^{-1}(k)D^0}{1+C(k,z)}}
\label{fkz2}
\end{equation}
and
\begin{equation}
F_S(k,z) = \frac{1}{z+\frac{k^{2}D^0}{1+C_S(k,z)}}, \label{fskz2}
\end{equation}
where $C(k,z)$ and $C_S(k,z)$ are the corresponding collective and
self memory functions  \cite{atomic2}.

As in the description for colloidal liquids \cite{scgle0},
$C(k,z)$ and $C_S(k,z)$ can be written in terms of the higher-order
memory functions $L_{UU}(k,z)$ and $L_{UU}^{(S)}(k,z)$, which
are the time-dependent correlation function of the configurational
component of the stress tensor \cite{atomic2}. The inclusion of such a
higher-order memory functions is only necessary if an accurate
description of the short-time dynamics becomes a priority. Our
present interest, however, refers primarily to the opposite time
regime, i.e., the long-time dynamics, which becomes relevant, for example, 
for the phenomenology of the glass transition, in which case we may refer only
to the primary memory functions $C(k,t)$ and $C_S(k,t)$.

The explicit comparison of the resulting equations for $F(k,z)$ and $F_S(k,z)$ 
for atomic liquids with those of a colloidal liquid (Eqs.(4.24) and (4.33)
of Ref. \cite{scgle0}) reveals the remarkable formal identity
between the long-time expressions for $F(k,t)$ and $F_S(k,t)$ of
an atomic liquid and the corresponding results for the equivalent
colloidal system. Analogously, to accurately describe the transition
from the ballistic regime to the diffusive behavior, the fundamental
difference between both types of liquids turns out to be found in the
definition of the diffusion coefficient $D^0$, given by
Eq. (\ref{dkinetictheory}) in the case of atomic liquids.

\subsection{Self-consistent description of the
\emph{long-time} dynamics of atomic liquids}\label{subsectionII.3}

The formal identity with the dynamics of Brownian liquids,
theoretically predicted in Refs. \cite{atomic1,atomic2},
immediately suggests that the long-time dynamic properties
of an atomic liquid will then coincide with the corresponding
properties of a colloidal system with the same $S(k)$,
provided that the time is scaled as $D^0t$ with the respective
meaning and definition of $D^0$. As mentioned before, this expectation
has been nicely confirmed by performing both Brownian dynamics (BD)
and Molecular dynamics (MD) simulations on the same prescribed
model system, and then, comparing the results of both simulations
with the appropriate reescaling, see, e.g., Fig. 1 of Ref. \cite{atomic1}
and Fig. 2(a) of Ref. \cite{atomic2}.

Thus, Eqs. (\ref{dzdt0p}),  (\ref{fkz2}), and (\ref{fskz2}) provide the
basis of a theory for the long-time dynamics of atomic liquids. The
simplest approximation to close this set of equations is exactly the same
employed in the SCGLE theory of colloid dynamics \cite{scgle1}. Hence, since
Eq. (\ref{dzdt0p}) writes $\Delta \zeta ^*(t)$ in terms of $F(k,z)$ and $F_S(k,z)$,
whereas Eqs. (\ref{fkz2}) and (\ref{fskz2}) describe $F(k,z)$, and $F_S(k,z)$
in terms of the corresponding memory functions $C(k,t)$ and $C_S(k,t)$, what
one needs to define a closed system of equations is independent expressions
for such memory functions.
The first major approximation thus introduced is a Vineyard-like approximation,
which consists in assuming the simplest connection between $C(k,z)$ and
$C_S(k,z)$, namely \cite{todos2},
\begin{equation}
C(k,z) = C_S(k,z). \label{vineyardlike}
\end{equation}
The second major approximation consists in interpolating $C_S(k,z)$
between its two exact limits at small and large wave-vectors by
means of a completely empirical interpolating function $\lambda(k)$,
chosen such that $\lambda(k\to 0) =1$ and $\lambda(k\to \infty)=0$.
The large wave-vector limit, $C_S(k\to \infty,t)$, is a
rapidly-decaying function of time, and hence we approximate it by its
vanishing long-time value. Thus, one can write $C_S(k,t)$ simply as
$C_S(k,t) = \lambda(k)C_S(k=0,t)$, or as
\begin{equation}
C_S(k,z) = \lambda(k)\Delta \zeta ^*(t), \label{interpolation}
\end{equation}
since one can demonstrate that $C_S(k=0,t)=\Delta \zeta ^*(t)$.

If one now incorporates both Vineyard-like and interpolation approximations
in Eqs. (\ref{fkz2}) and (\ref{fskz2}), one can then rewrite such equations as,
\begin{equation}
F(k,z) = \frac{S(k)}{z+\frac{k^{2}D^0S^{-1}(k)}{1+\lambda(k)
\Delta \zeta^*(t)}},
\label{fkz3}
\end{equation}
and
\begin{equation}
F_S(k,z) = \frac{1}{z+\frac{k^{2}D^0}{1+\lambda(k)
\Delta \zeta^*(t)}}, \label{fskz3}
\end{equation}
with $D^0$ given by the kinetic value provided in Eq.
(\ref{dkinetictheory}). This set of equations together with Eq.
(\ref{dzdt0p}) for $\Delta \zeta ^*(t)$ constitute the closed
system of equations that define the self-consistent theory of
the long-time dynamics of an atomic liquid. For the interpolating
function, the same functional form as in the colloidal case is
adopted, namely,
\begin{equation}
\lambda(k) \equiv \frac{1}{1+\left(\frac{k}{k_c}\right)^2},
\label{lambdadk}
\end{equation}
with $k_c$ being an empirically chosen cutoff wavevector,
sometimes related with the position of the maximum $k_{max}$ of
$S(k)$ by $k_c=a k_{max}$, with $a>0$ being the only
free parameter, determined by a calibration procedure
\cite{gabriel}.

In summary, Eqs. (\ref{dzdt0p}),  (\ref{dkinetictheory}), and
(\ref{fkz3})-(\ref{lambdadk}), constitute a self-consistent system
of equations for the time-dependent friction function
$\Delta \zeta^* (t)$ and the correlation functions
 $F(k,t)$ and $F_S(k,t)$ of an atomic liquid in the long-time
(or overdamped) regime. These equations provide an approximate
first-principles prediction of the \emph{diffusive} dynamics of
an atomic liquid, i.e., outside the short-time or ballistic regime.
They are, however, formally identical to the SCGLE theory of colloid
dynamics \cite{scgle1}, and hence, they express the aforementioned
dynamical equivalence between atomic and colloidal systems discussed
in detail in Refs. \cite{atomic1,atomic2,atomic3}. This predicted
long-time dynamical equivalence is best illustrated in terms of the mean
squared displacement (MSD), $W(t)\equiv \ (1/6) <(\Delta \textbf{R}(t))^2>$,
which in the case of Brownian liquids is the solution of the following equation,
\begin{equation}
W(t) =D^0t-\int_0^t \Delta \zeta^*
(t-t') W(t)dt',   \label{wdtintdifecoverdamped}
\end{equation}
and whose long-time limit provides a master curve for both atomic and
Brownian liquids, as explicitly illustrated in section \ref{sectionIII}.

\section{Crossover from  ballistic to diffusive dynamics}\label{sectionIII}

The self-consistent set of equations proposed in the
previous section is sufficient to describe the long-time dynamics
of an atomic liquid. Let us now complement this dynamic framework 
with additional approximations to allow the incorporation of the
ballistic short-time behavior. There are, of course, more than one 
concrete manners to carry out this task, and here, we propose
one based on the criterion of analytic and numerical simplicity, but also on
physical self-consistency. Although the ultimate interest in
this paper refers to the description of multicomponent atomic liquids,
for the sake of the discussion we shall address this issuue first in the
context of monocomponent systems. The extension to mixtures will be discussed
in the following section.

\subsection{\emph{Monocomponent} atomic SCGLE theory}\label{sectionIII.1}

To incorporate the correct short-time limit in the  framework of
the atomic SCGLE theory, our strategy consist in to obtain first
an accurate description of the crossover from  ballistic to diffusive
dynamics, well-known and typically observed in the MSD, $W(t)$. To this end,
let us notice that the exact equation for $W(t)$ which can be directly
derived from Eq. (\ref{eq6p}), reads
\begin{equation}
\tau^0\frac{dW (t)}{d t} +W(t) =D^0t-\int_0^t \Delta \zeta^*
(t-t') W(t')dt',  \label{wdtintdifec}
\end{equation}
where $\tau^0\equiv M/\zeta^0 = D^0/v_0^2$. This exact equation
can be now closed with the overdamped approximation for the friction
function $\Delta \zeta^* (t)$ that results from the solution of
Eqs. (\ref{dzdt0p}),  (\ref{dkinetictheory}), and (\ref{fkz3})-(\ref{lambdadk}).
Even though this approximation is only valid for $t>>\tau_0$,
at short times, the effects of the interparticle interactions on $W(t)$
(embodied in $\Delta \zeta^* (t)$)  are negligible. Nevertheless, the correct
short-time ballistic behavior of $W(t)$ is guaranteed by the presence of the first
(``inertial'') term of Eq. (\ref{wdtintdifec}). Thus, the solution
of this equation for $W(t)$ will provide an accurate interpolation
between its ballistic (short-time) and long-time (diffusive)
regimes, described respectively by the following asymptotic limits,
\begin{equation}
W(t)  \approx \frac{1}{2} v_0^2 t^2, \text{ \ \ for \ \ } t\to 0,
\label{wdt0short}
\end{equation}
where $v_0^2\equiv k_BT/M$, and
\begin{equation}
W(t)  \approx D_Lt, \text{ \ \ for \ \ }t\to \infty,
\label{wdt0long}
\end{equation}
with $D_L$ being the long-time self-diffusion coefficient,
determined by  the Green-Kubo relation
$D_L=\int_0^\infty dt V(t) = V(z=0)$ and by the velocity
autocorrelation function $V(t)$, which is obtained from Eq.
(\ref{eq6p}) in terms of  $\Delta \zeta (t)$, leading to
\begin{equation}
D_L=\frac{D^0}{1+ \Delta \zeta ^*} ,
\label{dlong}
\end{equation}
with
\begin{equation}
\Delta \zeta ^* \equiv \int_0^\infty \left[\frac{\Delta \zeta
(t)}{\zeta^0}\right]dt.  \label{deltazetastar}
\end{equation}

The most natural analogous procedure to incorporate the
ballistic regime in the ISFs would be to start with the exact
expressions provided in Eqs. (32) and (33) of Ref. \cite{atomic2},
in order to introduce general closure relations for the respective
memory functions to construct a self-consistent squeme.
In fact, we have followed such kind of procedure and found that the
structure of the resulting equations, which naturally should describe
the crossover from the ballistic and diffusive limits, sometimes
introduced a spurious oscillatory time dependence of the ISFs. This was
due to the fact that the analytic structure of such equations is actually
not consistent with the Gaussian limit of the ISFs, which is exact at
short-times and low particle densities. Although we might attempt to carry
out a formal derivation to sort out these difficulties, we have found
a simplified alternative consisting in start from the overdamped
approximations for the ISFs in Eqs. (\ref{fkz3}) and (\ref{fskz3})) to
interpolate between the two (short and long time) limits of these dynamical
properties.

Thus, let us recall now that the well-known Gaussian approximation for $F_S(k,t)$
and $F(k,t)$ is defined as \cite{boonyip}
\begin{equation}
F_S(k,t)= e^{-k^2W(t)},
\label{fsgaussian}
\end{equation}
and
\begin{equation}
F(k,t)= S(k)e^{-k^2W(t)/S(k)}.
\label{fgaussian}
\end{equation}
As was illustrated in Fig. 2 of Ref. \cite{atomic2},
the validity of these approximations is restricted to the
short-time regime, $t \lesssim \tau_0$, whereas the solution
for Eqs. (\ref{fkz3}) and (\ref{fskz3}) (which from now on will
be labeled with a superscript $D$, i.e., $F^D(k,t)$ and $F^D_S(k,t)$,
to denote the solution at long-times)
correctly describe the value of the ISFs in the complementary time
domain, $t \gtrsim \tau_0$. Hence, as a simple description of the
ballistic-to-diffusive crossover of the ISFs, we may propose to
approximate $F(k,t)$ and $F_S(k,t)$ of an atomic liquid
as the following simple exponential interpolation between the
two aforementioned limits, namely,
\begin{eqnarray}
F(k,t)=F^D(k,t)+\{S(k)\exp[-k^2W(t)/S(k)]\nonumber\\
-F^D(k,t)\}
\exp[-t/\tau^0],\label{fktinterpolation}
\end{eqnarray}
and
\begin{equation}
F_S(k,t)= F^D_S(k,t)+\{\exp [-k^2W(t)] -F_S^D(k,t)\}\exp[-t/\tau^0].\label{fsktinterpolation}
\end{equation}

In summary, we have proposed an approximate, but general,
first-principles SCGLE theory for the dynamical properties
of a simple monocomponent atomic liquid. This theory is finally 
outlined by Eqs. (\ref{dzdt0p}), (\ref{fkz3})-(\ref{lambdadk}), (\ref{wdtintdifec})
and (\ref{fktinterpolation})-(\ref{fsktinterpolation}), which provide
a protocol to determine the dynamical features of an atomic liquid
starting from the intermolecular forces, represented by the pair interaction
potential $u(r)$. This protocol is now spelled out and illustrated with a
specific application.

\subsection{Atomic hard sphere liquid.}\label{subsectionV.2}

Once the pair potential  $u(r)$ is specified, the first step is
to determine the static structure factor $S(k)$ according to any
equilibrium (exact or approximate) method. For example, for the
atomic version of a liquid of  hard spheres (HS) of diameter $\sigma$
and number density $n$, we can approximate $S(k)$ by its analytic
Percus-Yevick-Verlet-Weis (PYVW) expression
\cite{percusyevick,verletweis}, which is highly accurate throughout
the stable liquid regime. Using this $S(k)$ as an input, Eqs.(\ref{dzdt0p})
and (\ref{fkz3})-(\ref{lambdadk}) are solved to determine $\Delta \zeta^*(t)$
and the overdamped ISFs $F^D(k,t)$ and $F^D_S(k,t)$. The resulting $\Delta \zeta^*(t)$  is
then employed in Eq. (\ref{wdtintdifec}), whose solution for $W(t)$
provides the input of the expressions in Eqs.
(\ref{fktinterpolation}) and (\ref{fsktinterpolation})
for $F(k,t)$ and $F_S(k,t)$, which now incorporate the correct short-time
ballistic limit.

\begin{figure}[h]
\centering
\subfigure{ \label{fig:1a}
\includegraphics[scale=.15]{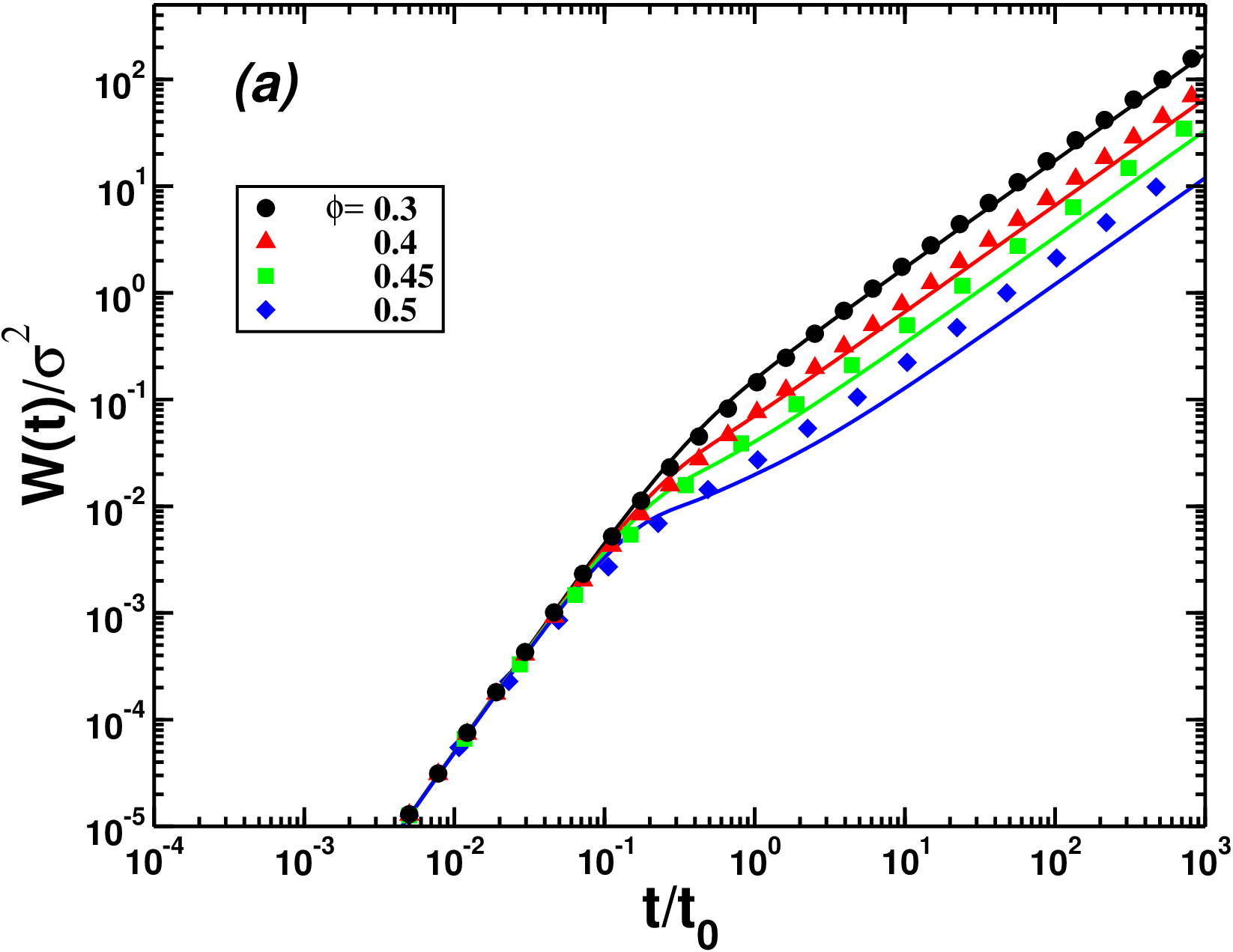}}
\subfigure{ \label{fig:1b}
\includegraphics[scale=.15]{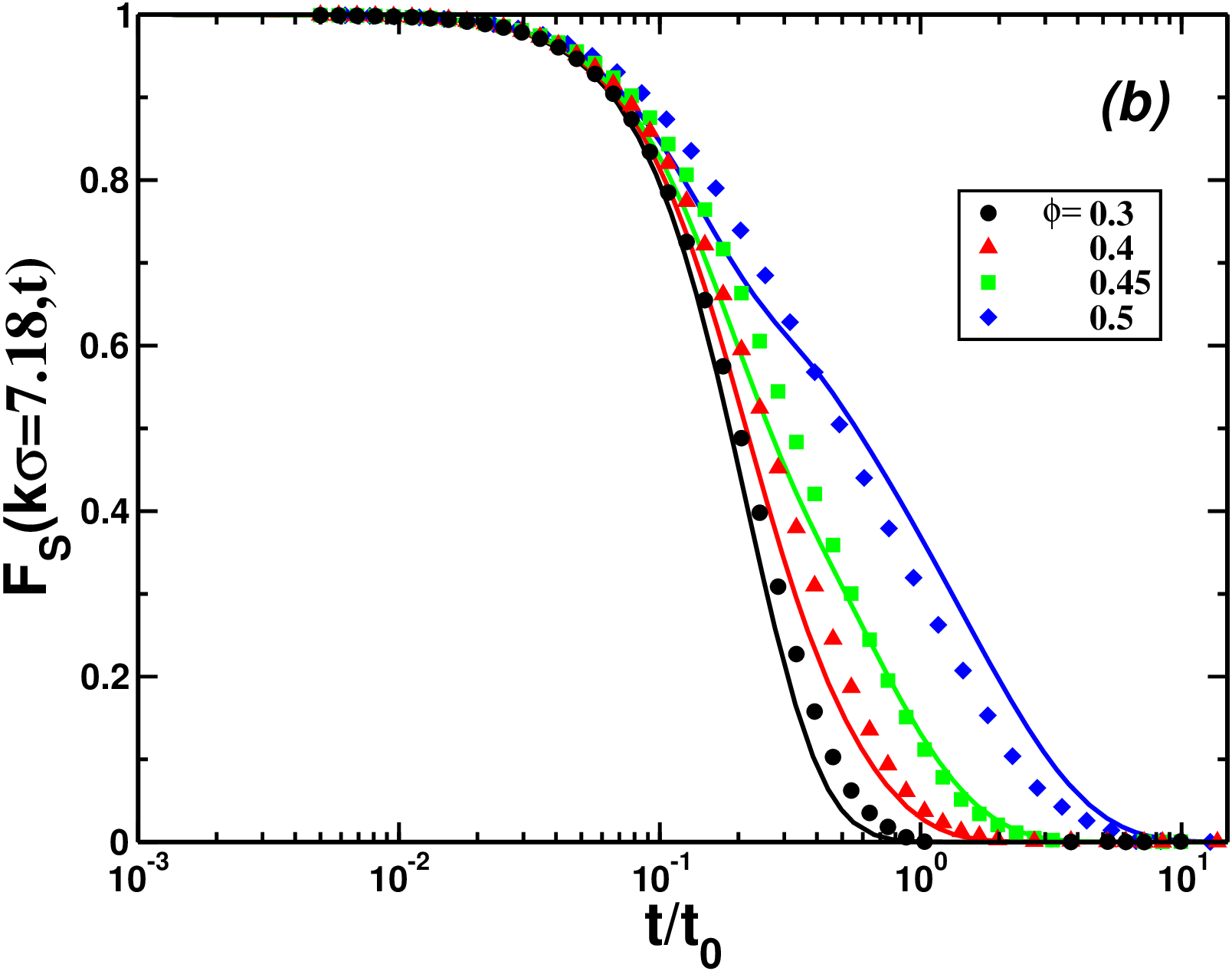}}
\subfigure{ \label{fig:1c}
\includegraphics[scale=.15]{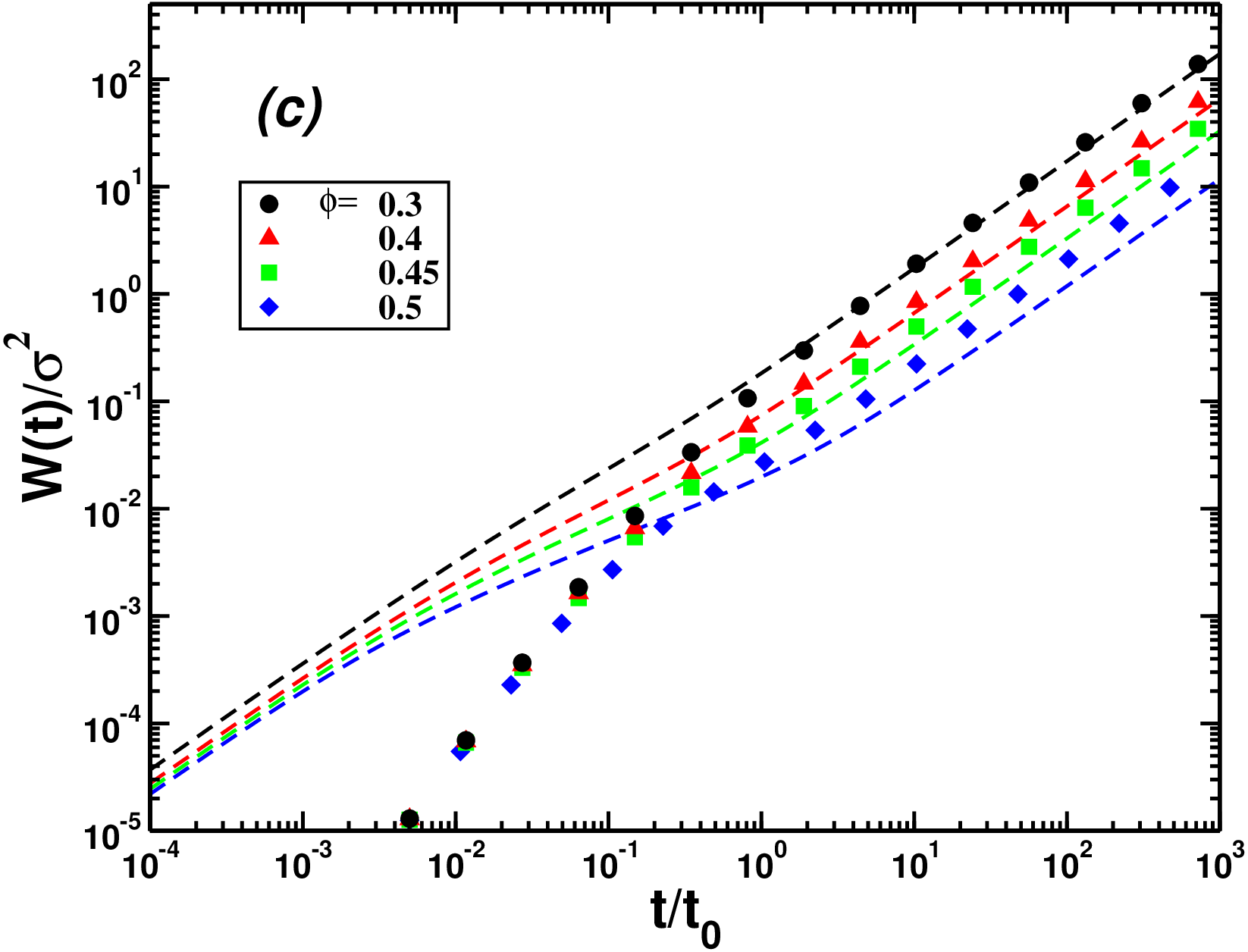}}
\subfigure{ \label{fig:1d}
\includegraphics[scale=.15]{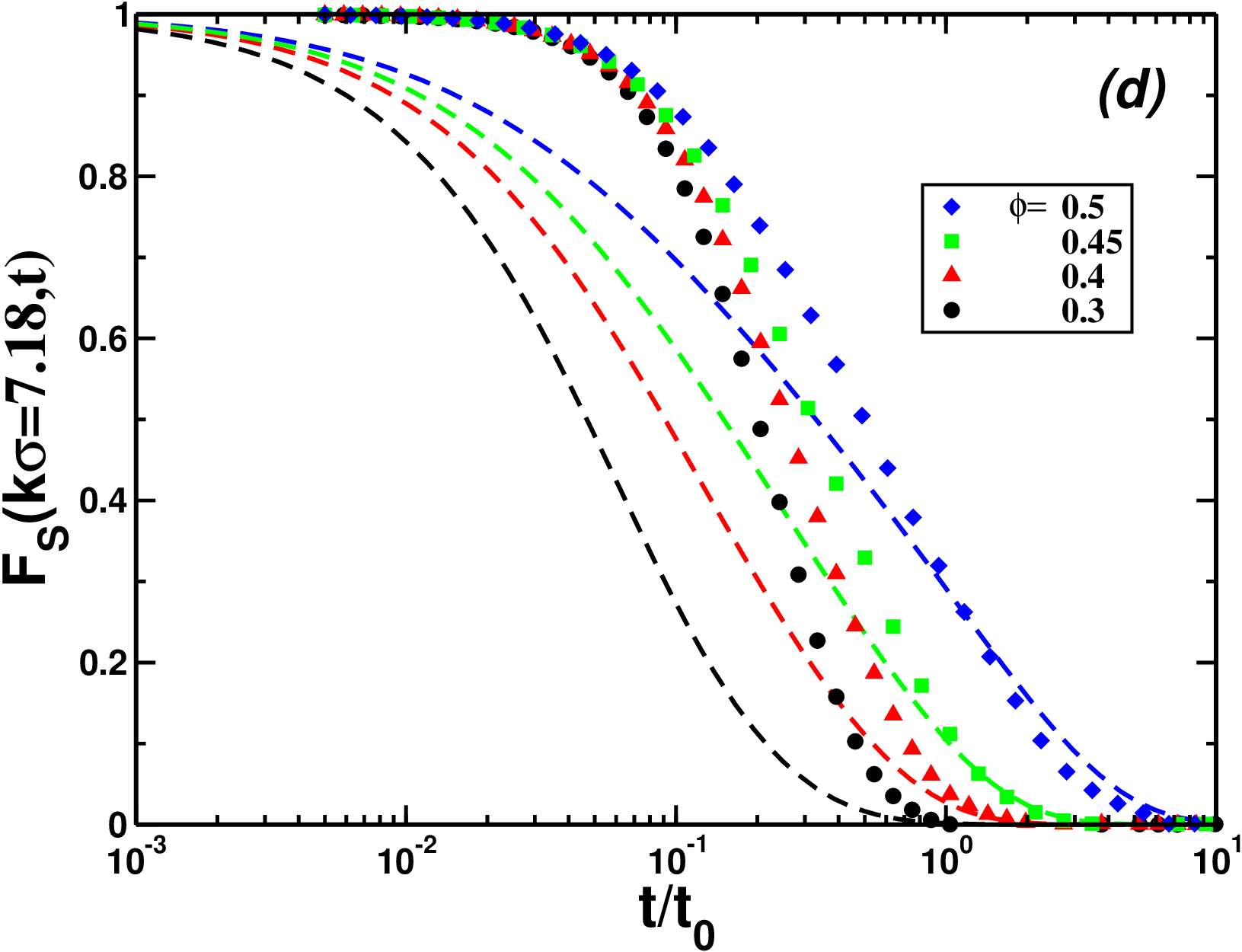}}
\caption{Mean-square displacement $W(t)$ (in units of $\sigma^2$)
and self-intermediate scattering function $F_S(k,t)$ of a hard sphere 
fluid as a function of the time $t$ (in ``molecular units''
$t_0=\sqrt{M\sigma^2/k_BT}$) for volume fractions $\phi=0.3, 0.4, 0.45$ and $0.5$.
In all cases, symbols are the results of our MD simulations. The solid lines in
(a) and (b) represent, respectively, the solution of Eq.
(\ref{wdtintdifec}) for  $W(t)$ and Eq. (\ref{fsktinterpolation})
for $F_S(k,t)$ at fixed wave vector $k\sigma=7.18$.
The dashed lines in (c) and (d) are the solution of, respectively, Eqs.
(\ref{wdtintdifecoverdamped}) and
(\ref{fskz3}).}
\label{fig1}
\end{figure}

The results for $W(t)$ and $F_S(k,t)$ for the atomic HS liquid are represented 
by the solid curves in Figs. \ref{fig:1a}
and \ref{fig:1b}, respectively. These results are plotted in terms
of the MD ``natural'' (lenght and time) units,  $\sigma$
and $t_0\equiv \sqrt{M\sigma^2/k_BT}$ at four values of the 
volume fraction $\phi=\pi n\sigma^3/6$, representative of the stable liquid
regime, namely, $\phi=0.3, 0.4, 0.45$ and $0.5$. In the same figures,
we have also included the corresponding data (solid symbols)
obtained by MD simulations. The details of the simulations are
provided in the following section. The comparison of Fig. \ref{fig:1a}
illustrates how, in spite of the fact that we have used the
solution $\Delta \zeta^*(t)$ of the \emph{overdamped} SGCLE theory
to solve Eq. (\ref{wdtintdifec}), the presence of the inertial term
in the same equation ensures that the resulting MSD has the correct
short-time limit $W(t) \approx v_0^2 t^2/2$, and correctly describes
the passage of $W(t)$ from this ballistic regime to its long-time diffusive
limit $W(t) \approx D_Lt$. We observe that for volume fractions
smaller than $0.5$, the quality of the agreement is better than that
observed at the volume fraction $0.5$. For volume fractions in the
metastable regime, $\phi\gtrsim 0.5$, these quantitative differences becomes
larger (data not shown). In addition, deviations are also observed at
intermediate times, which amplify slightly at the larger volume fractions.
These inaccuracies, however, may be perfectly tolerable in a theory
with no adjustable parameters.

For completeness, in Fig. \ref{fig:1c} the solution of the
\emph{overdamped} version of Eq. (\ref{wdtintdifec}) (see
Eq. (\ref{wdtintdifecoverdamped})) using the same \emph{overdamped}
$\Delta \zeta^*(t)$ as input has been included. The resulting MSD,
represented by the dashed lines, describes the dynamics of the
corresponding Brownian liquid whose solution satisfies the
short- and long-time limits, i.e.,
$W(t\to 0) \approx D^0t$ and  $W(t\to \infty) \approx D_L t$,
respectively. The comparison between the theoretical curves and the
simulation data demonstrates the theoretically predicted long-time
dynamic scaling between atomic and colloidal systems in the context of
the MSD, as already reported in Ref. \cite{atomic3}.

In Fig. \ref{fig:1b} the theoretical predictions for $F_S(k,t)$ are compared
with the corresponding MD results at the same volume fractions. This
comparison indicates that the simple approximation in
Eq. (\ref{fsktinterpolation}) provides a very good quantitative representation
of $F_S(k,t)$ at very low volume fractions, although the inaccuracies of the
theoretical results for $W(t)$ shown in Fig. \ref{fig:1a} manifest themselves
in the differences observed, particularly at $\phi=$ 0.5 at intermediate times.

In Fig. \ref{fig:1d}, the quantitative differences
between the  solution of the \emph{overdamped} SCGLE equations
(dashed lines) for $F_S(k,t)$ and the MD simulation data are illustrated.
From this comparison, which complements the information provided in Fig. \ref{fig:1b},
one can see that the final decay of $F_S(k,t)$ is only well captured
for $\phi \geq 0.45$. Appreciable differences are observed, however,
at shorter times ($t \lesssim  t_0$), derived from the fact that in atomic
systems the relaxation of $F_S(k,t)$ in this time
regime reflects the early ballistic displacements of the particles,
not captured within the \emph{overdamped} SCGLE theory.

Thus, one can conclude that the overdamped SCGLE equations provide a good
representation of the hard-sphere self-ISF at times longer
than $t_0$, in particular, for volume fractions beyond a threshold value
around $\phi \geq 0.45$. Below this threshold value, the overdamped SCGLE
theory fails to capture the essentially ballistic decay of the ISF
$F_S(k,t)$ of atomic liquids. This threshold coincides
approximately with the freezing volume fraction, and is actually
the same as the crossover volume fraction referred to in Ref.
\cite{atomic2}, above which the dynamical equivalence between atomic
and Brownian liquids is also exhibited by $F_S(k,t)$.
The fact that this dynamical equivalence holds better when increasing
the volume fractions in the metastable regime makes it specially
relevant to understand the common phenomenology of colloidal and
atomic glass formers, a point that will be addressed elsewhere \cite{todoss}

\section{SCGLE formalism for \emph{multicomponent} atomic liquids}\label{sectionIV}

We now describe the multicomponent extension of the SCGLE theory for atomic
liquids. Since such an extension is in reality rather
straightforward, we shall not go in any detail through each of the arguments
reviewed in the monocomponent case. Instead, we only summarize the
resulting set of SCGLE equations that describe the dynamics of a
\emph{multicomponent} atomic liquid and provide two illustrative
applications involving the direct comparison with the corresponding MD
simulation results. In this section, we also briefly describe the simulation
methods employed.

\subsection{Dynamics of multicomponent atomic liquids}\label{subsectionIV}

Let us consider now an atomic liquid at temperature $T$
inside a volume $V$ and composed of $N$ spherical particles
belongin to $\nu$ different species (labeld by the index
$i=1,2,...,\nu$). Thus, $N=\sum_{i=1}^{\nu}N_{i}$,
where $N_{i}$ is the number of particles of the species $i$,
each of one having a mass $M_i$ and a diameter $\sigma_i$.
One can also define the number concentration of species $i$ as
$n_{i}=N_{i}/V$. The relevant structural information of these
multicomponent atomic liquid is contained in the elements  $S_{ij}(k)$
of the $\nu \times \nu$ matrix $S(k)$ of partial static structure factors,
whereas the most fundamental dynamical information is contained in the
$\nu\times\nu$ matrices $F^D(k,t)$ and $F^D_S(k,t)$, whose elements are the
collective  and partial self-intermediate scattering functions $F^D_{ij}(k,t)$
and $\delta _{ij}F^D_{S,i}(k,t)$, respectively.

As explained in detail in Refs. \cite{marco2,rigo1}, within the SCGLE formalism
the time-evolution equations for the matrices $F^D(k,t)$ and $F^D_S(k,t)$, written
in Laplace space, reads
\begin{eqnarray}\label{fluct5ppp}
F^D(k,z)= \{zI + k^2 D \cdot [zI+  \lambda(k)
\cdot  \Delta \hat \zeta^*(z) ]^{-1} \nonumber\\
\cdot S^{-1}(k) \}^{-1} \cdot S(k),
\end{eqnarray}
and
\begin{eqnarray}\label{fluct5sppp}
F^D_S(k,z) = \{zI + k^2 D\cdot  [zI+ \lambda(k)
\cdot \Delta \hat \zeta^*(z) ]^{-1} \}^{-1},
\end{eqnarray}
where $D$ and $\lambda(k)$ are diagonal matrices given by
$D_{ij}\equiv\delta_{ij}D^0_{i}$ and
$\lambda _{ij}(k)=\delta_{ij}[1+( k/k^{c}_i)]^{-1}$
and with $D^0_{i}$ being the short-time self-diffusion
coefficient of species $i$, which depends on the
masses $M_i$, the temperature, and the size of the particles,
in a manner that extends the monocomponent kinetic expression
in Eq. (\ref{dkinetictheory}). The parameter $k^{c}_i$ is a
empirical cut-off wave-vector written as $k^{c}_i=a k^{max}_i$,
in which $k^{max}_i$ is the position of the maximum of
$S_{ii}(k)$ and $a>0$ is again the only free parameter,
eventually determined by a calibration procedure \cite{gabriel}.

The  $i$th diagonal element $\Delta \zeta^*_{i}(z)$
of the matrix $\Delta \hat \zeta^*(z)$ is the time-dependent friction
function of particles of species $i$, and according to Refs.
\cite{marco2,rigo1} is given by
\begin{eqnarray}
\Delta \zeta^*_{i} (t)=\frac{D^0_i}{3( 2\pi
) ^{3}}\int d {\bf k}\ k^2 [F^D_S(t)]_{ii}[h
\cdot \sqrt{n}  \cdot  S^{-1} \nonumber \\
\cdot F^D(t) \cdot S^{-1} \cdot   \sqrt{n}\cdot h
]_{ii}, \label{dzdtppp}
\end{eqnarray}
with the elements of the $k$-dependent matrix $h$ given by
$ h =\sqrt{n}^{-1} \cdot (S-I) \cdot
\sqrt{n}^{-1}$, where the elements
of the matrix $\sqrt{n}$ are $[\sqrt{n} \ ]_{ij}\equiv
\delta_{ij} \sqrt{n_{i}}$, and
we have systematically omitted the argument $k$ of the
$\nu\times \nu$ matrices $h(k)$, $S(k)$, $F^D(k,t)$, and $F^D_S(k,t)$.

Let us now write the corresponding results
for the mean-square displacement, $W(t)$, with its
diagonal elements, $W_{i}(t)$, i.e., the MSD
of particles of species $i$, given by

\begin{equation}
\tau_i^0\frac{dW_i(t)}{dt} +W_i(t) =D_i^0t-\int_0^\tau
\Delta \zeta^*_i (t-t') W_i(t')dt', \label{modwdtintdifec}
\end{equation}
where $\tau_i^0\equiv M_i/\zeta^0_i$,
and with  $\zeta^0_i=k_BT/D_i^0$. Finally, the exponential
interpolation functions for the $\nu\times\nu$ matrices $F(k,t)$ and
$F_S(k,t)$ of a multicomponent atomic liquid can be written as,

\begin{eqnarray}\label{fcmix}
F(k,t)= F^D(k,t)+\{ S(k) \cdot\exp [-k^2W(t)\cdot S^{-1}(k)] \nonumber\\
-F^D(k,t)\}\exp[-t/\tau_i^0],
\end{eqnarray}
and
\begin{eqnarray}\label{fsmix}
F_S(k,t)= F^D_S(k,t)+\{\exp [-k^2W(t)] - \nonumber\\
F_S^D(k,t)\}\exp[-t/\tau_i^0].
\end{eqnarray}

Eqs. (\ref{fluct5ppp})-(\ref{fsmix}) thus extend to mixtures
the SCGLE formalism for an atomic liquid. In what follows, its use will be
illustrated with two examples involving binary hard-sphere
mixtures. The self-consistent solution of these equations
requires the previous determination of the matrix $S(k)$
of partial static factors, which will be obtained using again
the Percus-Yevick approximation for multicomponent
\cite{baxterpymixture} liquids and with the corresponding
Verlet-Weiss correction ($\phi \to \phi-\phi^2/16$)
\cite{verletweis,williamsvanmegen}. Once $S(k)$ has been
determined, we solve  Eqs. (\ref{fluct5ppp})-(\ref{dzdtppp})
to obtain the matrices $\Delta \zeta^*(t)$, $F^D(k,t)$, and
$F^D_S(k,t)$, which describe the dynamics of the multicomponent
atomic liquid in the diffusive or overdamped regime. We then
use these results in Eqs. (\ref{modwdtintdifec})-(\ref{fsmix})
to determine the MSD, $W(t)$, and the ISFs $F(k,t)$ and
$F_S(k,t)$, which include the correct short-time ballistic
regime. 

\subsection{Molecular dynamic simulations}\label{subsectionV.1}

As said in the introduction, to test this SCGLE theory of the 
dynamics of atomic mixtures we have carried out event-driven molecular 
dynamics simulations in the context of two illustrative applications: 
a polydisperse HS liquid, modeled as a moderately size-asymmetric binary 
HS mixture, and a genuine, highly size-asymmetric binary HS mixture. Let
us now briefly describe the simulation
methods employed in each of these applications.

To simulate a polydisperse monocomponent HS liquid we
followed the methodology explained in Ref. \cite{gabriel},
using \emph{event-driven} MD simulations. The simulations
were carried out with $N = 1000$ particles in a volume $V$.
The diameters of these particles are uniformly distributed between
$\overline{ \sigma} (1-w/2)$ and $\overline{ \sigma} (1+w/2)$,
with $\overline \sigma$ being the average particle diameter. We
have considered the case $w=0.3$, which corresponds to a polydispersity
$s_\sigma = w/\sqrt{12}=0.0866$. We have assumed that all particles
have the same mass $M$ and the results are displayed in reduced units,
therefore, $\overline{\sigma}$ and $\overline{\sigma}\sqrt{M/k_BT}$ are
used as units of length and time, respectively. To improve the statistics
and reduce the uncertainties, every correlation function is obtained
over the average of 10 independent realizations for each volume fraction
considered. The same protocol was employed to simulate the particular 
case of a monodisperse HS liquid ($w=0$). In general, the effect of 
polydispersity on the dynamic properties $W(t)$ and $F_S(k,t)$ is not
as dramatic as in the structure \cite{zaccareli}, and the differences 
are even less noticeable in the stable liquid regime illustrated in Fig. \ref{fig1},

In the case of highly asymmetric binary mixtures, we have
carried out \emph{event-driven} MD simulations for a HS binary
system with asymmetry parameter $\delta\equiv\sigma_s/\sigma_b=0.2$.
Here the labels \emph{s} and \emph{b} corresponds, respectively, to
\emph{small} and \emph{big}.
Thus, we have simulated a system of $N(=N_b+N_s)$ particles,
consisting in $N_b$ \emph{big} particles and $N_s$ \emph{small}
particles, in a volume \emph{V}.
More specifically, we have considered three different state points in
the control parameter space of the system spanned by the pair
($\phi_b,\phi_s$), where
$\phi_{i}=\pi n_{i}\sigma_{i}^3/6$ and
$n_{i}=N_{i}/N$ ($i=b,s$).
Given $N_b$, the size of the cubic simulation box was adjusted in order
to control $\phi_b$. Then, $N_s$ was adjusted to control $\phi_s$
(see Ref. \cite{todoss}).
The specific simulated state points were \textbf{I}$=(0.45,0.05)$, with
$N_b=200$ and $N_s=2778$; \textbf{II}{$=(0.45,0.2)$, with
$N_b=150$ and $N_s=8333$; and \textbf{III}$=(0.6,0.05)$, with
$N_b=200$ and $N_s=2083$. For the state points \textbf{I} and
\textbf{III}, $10$ realizations of the system were performed,
\emph{i.e.\/}, runs with 10 different \emph{seeds} have been used
to explore the available phase space and to improve the statistics.
For the state point \textbf{II}, 5 different
\emph{seeds} have been considered. In these simulations, the unit of length
is defined by the diameter of the
large particles, $\sigma_b$, and the unit of mass is defined as the mass of
the big particles, $M_b$. The mass densities,
$\rho^M_{i}\equiv M_{i}/v_{i}$
($v_{i}=4\pi\sigma^3_{i}/3, i=s,b$), are set equal to define
the mass of the small particles.
Setting $k_B=1$, the unit of time is defined from the equipartition theorem
$\langle v_{b}^2\rangle=3k_BT/M_{b}$.
Periodic boundary conditions were employed in all directions.
It is also worth to stress that, for the points \textbf{I} and \textbf{II},
we have used a \emph{waiting} time $t_w=10^3$, while for \textbf{III}
we let $t_w=10^4$, in order to avoid aging effects (see Ref. \cite{todoss}).
Finally, it should be mentioned that in order to generate non-overlapping initial
configurations, a soft-core standard molecular dynamics with a repulsive short-range
potential and decreasing temperature was implemented \cite{vargas}.
This soft-core MD starts from a completely random
initial configuration.

\subsection{Polydisperse hard-sphere atomic liquid}\label{subsectionV.3}

Let us now proceed to solve  the SCGLE Eqs. (\ref{fluct5ppp})--(\ref{fsmix}) 
for the first of the two examples just described, namely, a polydisperse HS 
liquid with uniform size distribution and polydispersity $s_\sigma =0.0866$. 
We can model this distribution with its discretize version  by  partitioning 
the interval $\overline{ \sigma} (1-w/2)\le \sigma \le \overline{ \sigma} (1+w/2)$ 
in $\nu$ equally sized bins, and treat the polydisperse liquid as a $\nu$-component 
mixture. To compare  with the simulations we then compute the total properties, such 
as  $W(t)\equiv \sum_{i=1}^\nu x_i W_i(t)$, $F(k;t)=\sum_{i,j=1}^\nu \sqrt{x_i x_j}F_{ij}(k,t)$, 
and $F_S(k,t)\equiv \sum_{i=1}^\nu x_i F^S_i(t)$, where $W_i(t)$, $F_{ij}(k,t)$, 
and $F^S_i(t)$ are the partial MSD, collective ISFs, and self ISFs of the mixture,
and where $x_i=n_i/\sum_{i=1}^\nu n_i$. 

To solve Eqs. (\ref{fluct5ppp})-(\ref{fsmix}) we need first to determine its 
static input, \emph{i.e.,} the partial static structure factors, but as said
above, for our HS system these will be provided by the multicomponent 
Percus-Yevick approximation with its Verlet-Weiss correction \cite{baxterpymixture,verletweis,williamsvanmegen}. 
We also need to previously determine the short-time self-diffusion coefficients 
$D^0_i$  ($i=1,2 ...\nu$). Unfortunately, the  random-flight arguments involved in
the derivation of the monocomponent kinetic-theoretical expression in 
Eq.  \ref{dkinetictheory} are not easily generalized to the case of an arbitrary 
$\nu$-component atomic liquid. Thus, if we do not wish to treat these as free
adjustable parameters, we must resort to additional approximations or simplifications,
as we do in the present application. 

Thus, to model the simulated HS polydisperse liquid, let us use the approach
just described in its simplest form, i.e., by considering an equimolar HS  binary mixture,
with components having number concentrations
$n_1=n_2=n_t/2$ and particle diameters $\sigma_1=\overline{\sigma}(1-\epsilon)$
and  $ \sigma_2=\overline{\sigma}(1+\epsilon)$, with $\epsilon=0.0866$ chosen such
that the mean diameter, $\bar{\sigma}$, and mean-square diameter
$\bar{\sigma^2}$ (and hence, the polydispersity), is the same as that of the simulated system.
Furthermore, given the small asymmetry between the constituent particles
($\delta\approx0.84$),  it is reasonable to approximate the short-time self-diffusion 
coefficients $D^0_1$ and $D^0_2$ as $D^0_1=D^0_2=D^0_{eff}$  where $D^0_{eff}$ is the
short-time self-diffusion coefficient of an effective monodisperse system with concentration
$n_t$ and diameter $\bar{\sigma}$, given by  (see Eq. \ref{dkinetictheory})
\begin{equation}
D^0_{eff}\equiv \frac{3}{8}\left(\frac{k_BT}{ \pi M}\right)^{1/2}\frac{1}{n_t\bar{\sigma}^2}.
\label{dkinetictheoryEFEC}
\end{equation}
As it is shown in what follows, this proposal allows us to
provide a good representation of the referred dynamical properties
of the simulated polydisperse fluid.

\begin{figure}[h]
\centering
\subfigure{ \label{fig:2a}
\includegraphics[scale=.27]{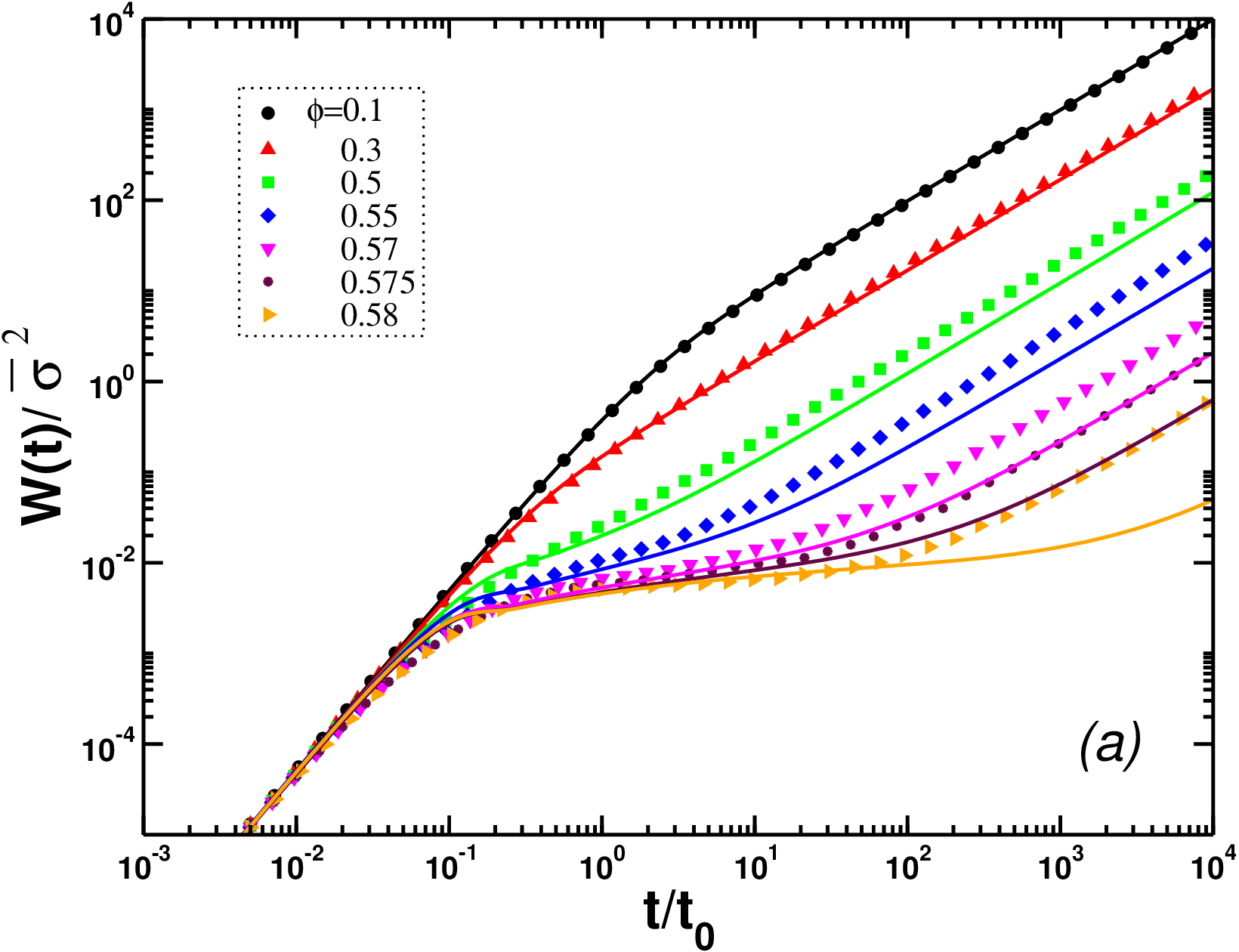}
}
\subfigure{ \label{fig:2b}
\includegraphics[scale=.27]{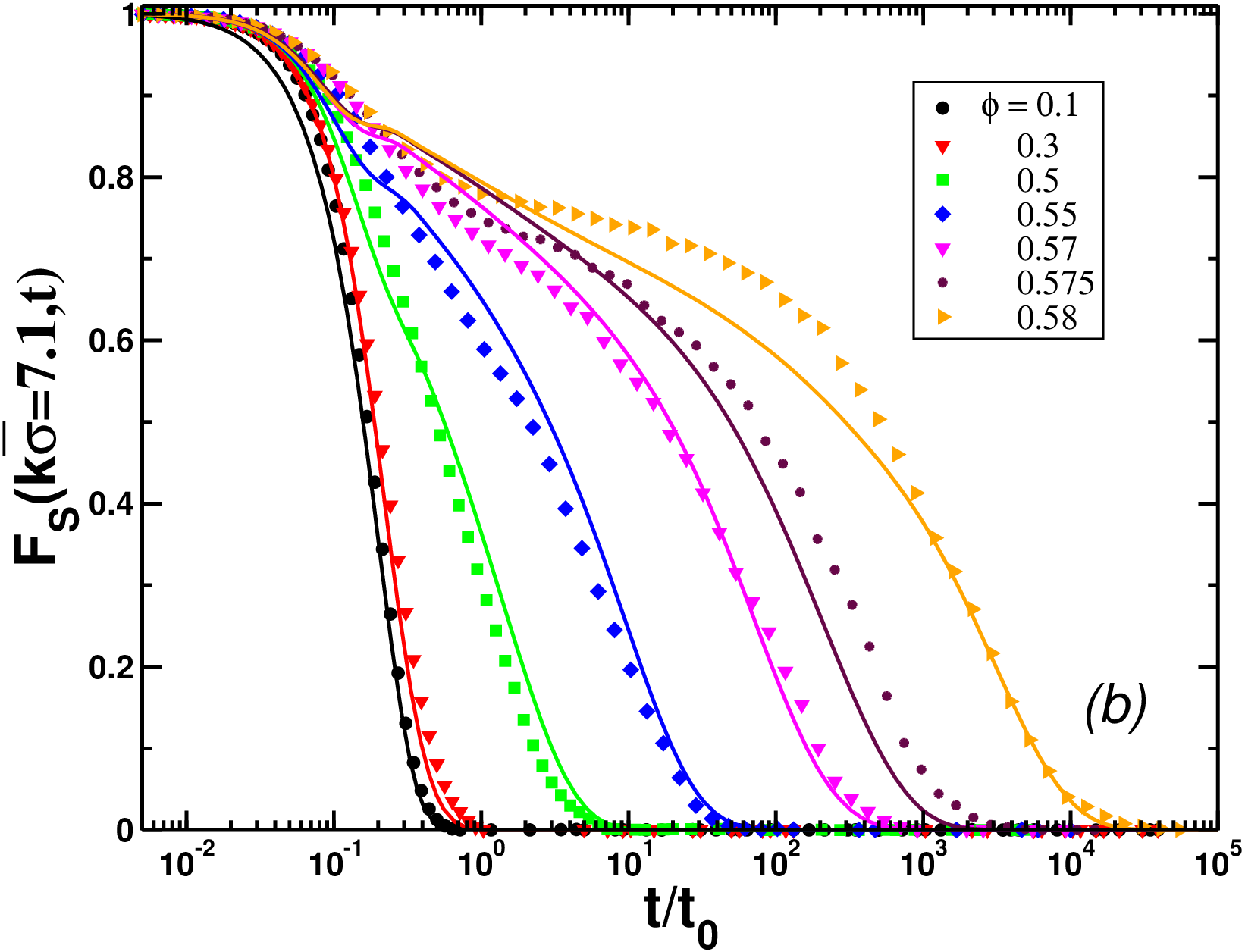}
}
\caption{ (a) Mean-square displacement $W(t)$ and (b) self-ISF,
$F_S(k,t;\phi)$, at fixed $k\bar{\sigma}=7{.}1$, as a
function of  time $t$ (in ``molecular" units $t_0\equiv \bar{\sigma}\sqrt{M/k_BT}$)
for the polydisperse hard-sphere atomic liquid (polydisersity = 8.66 \%) at the
indicated volume fractions. In both figures, solid symbols represents the MD
simulation data and the solid lines are the predictions of the SCGLE formalism
for atomic liquids.}
\label{fig2}
\end{figure}

In Fig. \ref{fig2} we show the theoretical predictions for $W(t)$ and
$F_S(k,t)$ compared against the corresponding MD simulation results for
a set of volume fractions. This figure  illustrates how, using the
\emph{overdamped} friction functions $\Delta \zeta^*_{i}(t)$ in the
time-evolution equations for
$W_{i}(t)$ (Eqs. (\ref{modwdtintdifec})) and assuming $D^0_1=D^0_2=D^0_{eff}$,
an effective MSD (solid lines) is generated, which nicely describes the ballistic regime, 
although some quantitative differences appears at long-times. Such
deviations, however, should be expected considering the calibration procedure employed to fix the only
phenomenological parameter of the SCGLE theory (the cutoff wave-vector
$k^c_i$) as $k^{c}_i=a k^{max}_i$, with $a=1.119$. This value was chosen 
in order to get the best \emph{overall} fit of the theory with the simulation data
for the so-called $\alpha$-relaxation time $\tau^{(\alpha)}$, defined by  
$F_S(k\bar{\sigma}=7.1,t=\tau^{(\alpha)})=e^{-1}$. For this reason, the \emph{overall} 
long-time agreement with the simulation results of
the predicted  $F_S(k,t)$ (see Fig. \ref{fig:2b}) is in general better than that of $W(t)$.
Similar to the MSD, one observes that for volume fractions smaller than 0.5, the quality
of the agreement at short and long times is better than that seen in the metastable regime,
$\phi\gtrsim 0.5$, where  there are some quantitative differences.
Thus, in spite of the use of the approximate effective diffusion coefficient in
Eq. \ref{dkinetictheoryEFEC}, the results show a nice agreement between simulation and
theory.

\subsection{Highly size-asymmetric binary mixture}\label{subsectionV.4}

The previous example illustrates a simple manner to apply the SCGLE
theory of atomic mixtures to represent the dynamical properties of a
polydisperse fluid. The small degree of polydispersity allowed us to 
use the approximation $D^0_1=D^0_2=D^0_{eff}$. Let us now turn our 
attention to the case of highly asymmetric HS binary mixtures, illustrated
with the simulated mixture with size asymmetry $\delta =0.2$ (and mass 
asymmetry $M_s/M_b=\delta ^3 = 0.008$). The theoretical procedure is exactly
the same, except that now we need to keep track of the partial dynamic properties
$W_i(t)$, $F_{ij}(k,t)$, and $F^S_i(t)$, and the size- and
mass-asymmetries are much larger. This forces to consider more accurate expressions 
for the short-time diffusion coefficients  $D^0_1$ and $D^0_2$ provided by kinetic 
theory, namely \cite{mcquarrie}, 
\begin{eqnarray}
D^0_b =\frac{3}{2}\left( \dfrac{ k_B T}{ \pi M_b}\right)^{1/2}
\left[ \dfrac{1}{4\sigma_b^2n_b+ (\sigma_b+\sigma_s)^2 n_s}\right]
\label{dkinetictheoryBIGG}
\end{eqnarray}
\begin{eqnarray}
D^0_s =\frac{3}{2}\left( \dfrac{k_B T}{ \pi M_s}\right)^{1/2}\left[\dfrac{1}{ (\sigma_b+\sigma_s)^2 n_b + 4\sigma_s^2n_s}\right],
\label{dkinetictheorySMALL}
\end{eqnarray} 
which contains the monocomponent expression in Eq.  (\ref{dkinetictheory}) as a particular case.

In practice, however, since the big particles are far more massive than the small particles, 
in determining the self-diffusion coefficient of the large spheres, 
we simply assume that the most representative collisions contributing to  $D^0_b$
are those among big particles, and thus, we may  approximate such coeficient by 
(see Eq. (\ref{dkinetictheory}))
\begin{equation}
D^0_b\equiv \frac{3}{8}\left(\frac{k_BT}{\pi M_b}\right)^{1/2}\frac{1}{n_b\sigma^2_{b}}.
\label{dkinetictheoryBIG}
\end{equation}
For the small particles, however, we do consider both types of
collisions, \emph{i.e.}, those involving only small particles and 
those among small and large ones, so that

\begin{eqnarray}
 \dfrac{D^0_s}{D^0_b} =4 \left(\dfrac{M_b}{M_s}\right)^{1/2}\left[\dfrac {\sigma_b^2n_b}{(\sigma_b+\sigma_s)^2 n_b + 4\sigma_s^2n_s}\right].
\label{dsmall}
\end{eqnarray} 
\begin{figure}[h]
\centering
\subfigure[]{ \label{fig:3a}
\includegraphics[scale=.149]{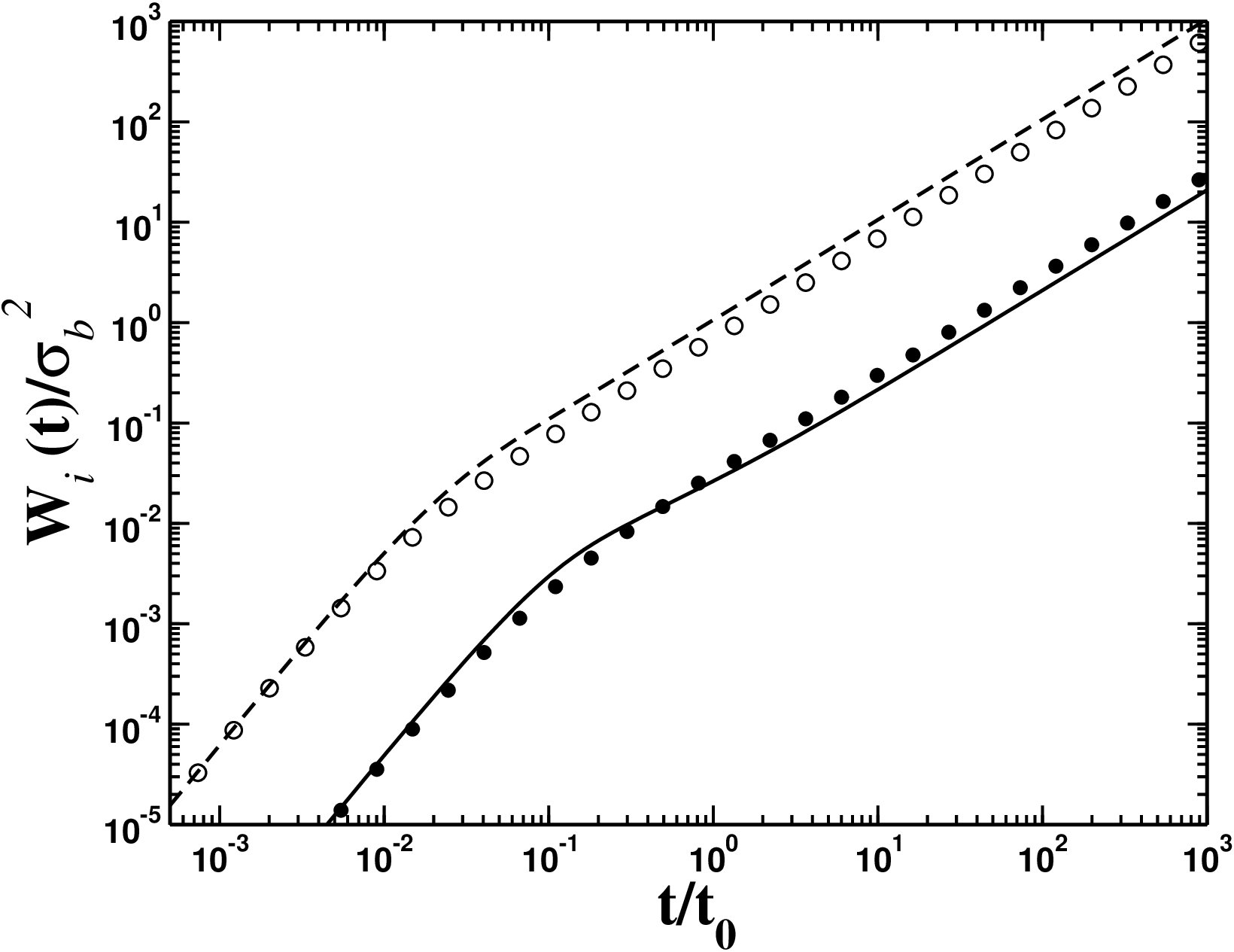}}
\subfigure[]{ \label{fig:3b}
\includegraphics[scale=.149]{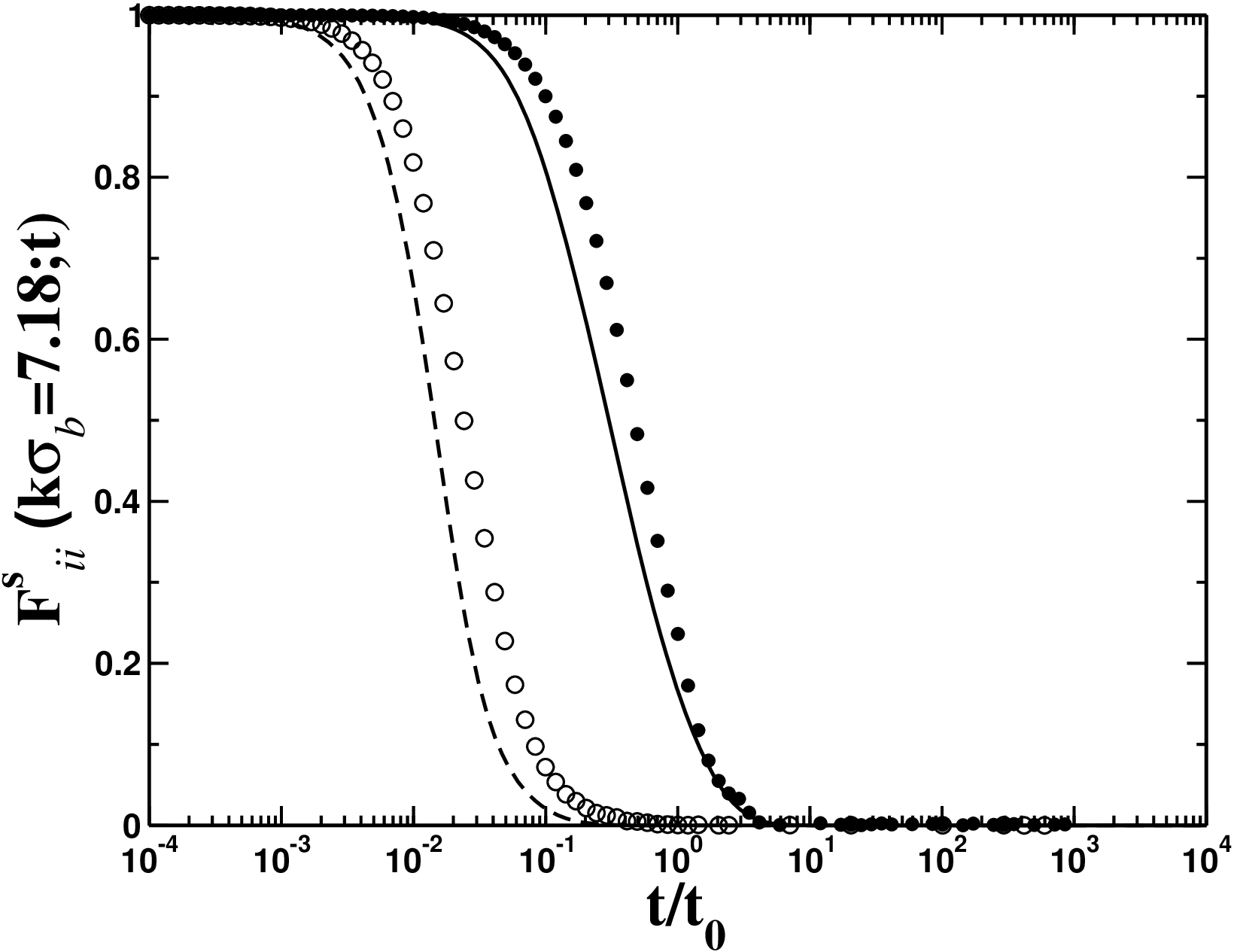}}
\subfigure[]{ \label{fig:3c}
\includegraphics[scale=.149]{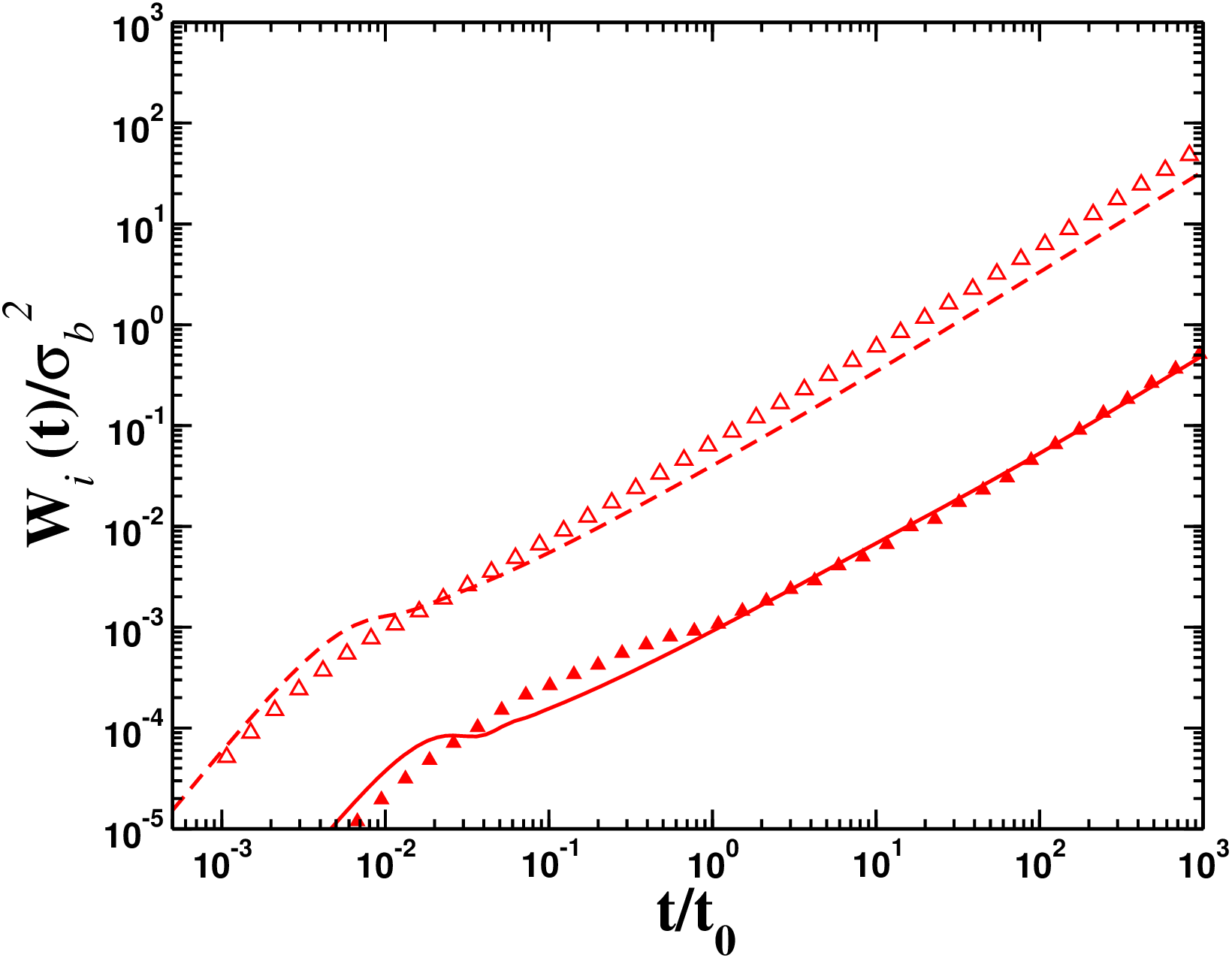}}
\subfigure[]{ \label{fig:3d}
\includegraphics[scale=.149]{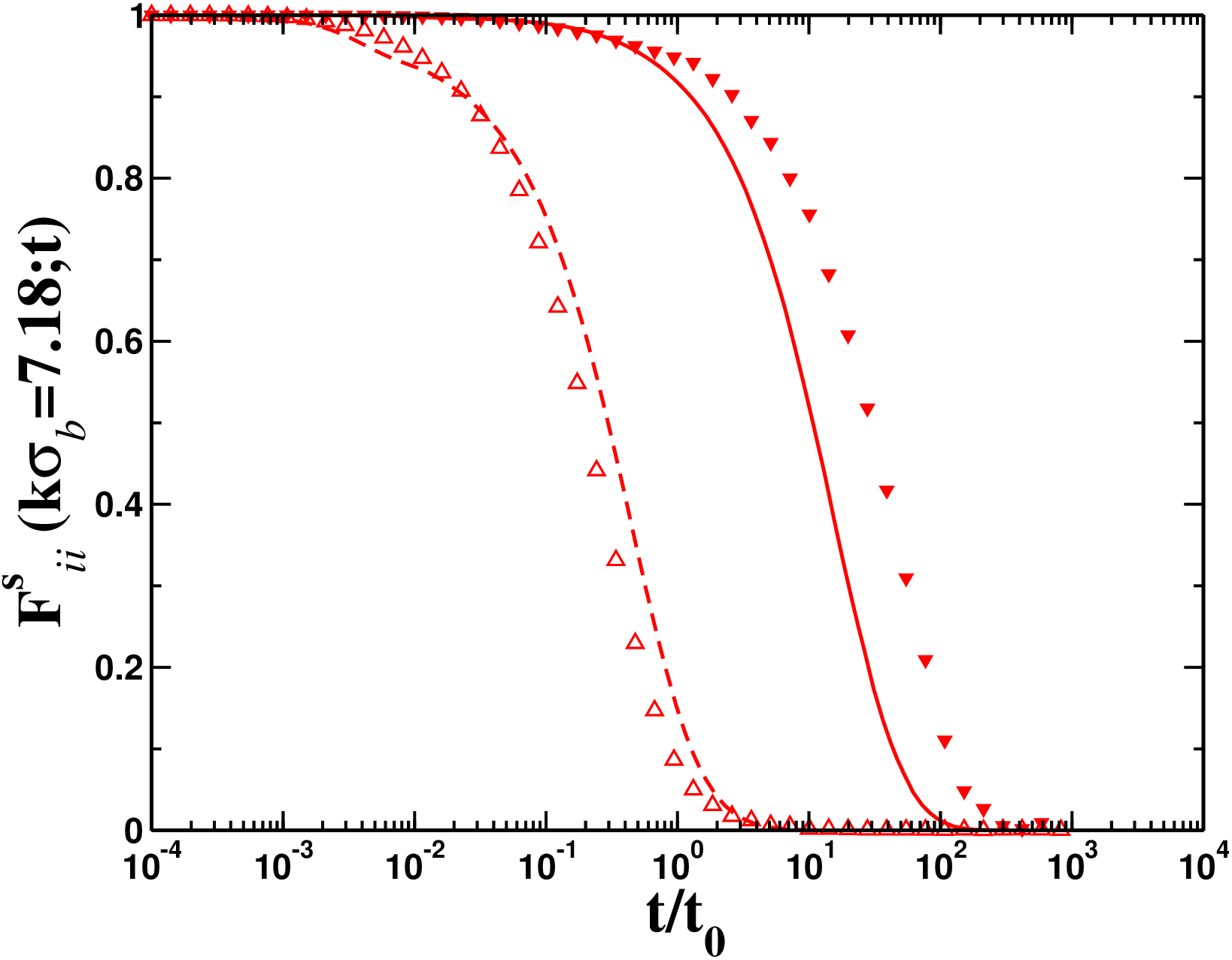}}
\subfigure[]{ \label{fig:3e}
\includegraphics[scale=.149]{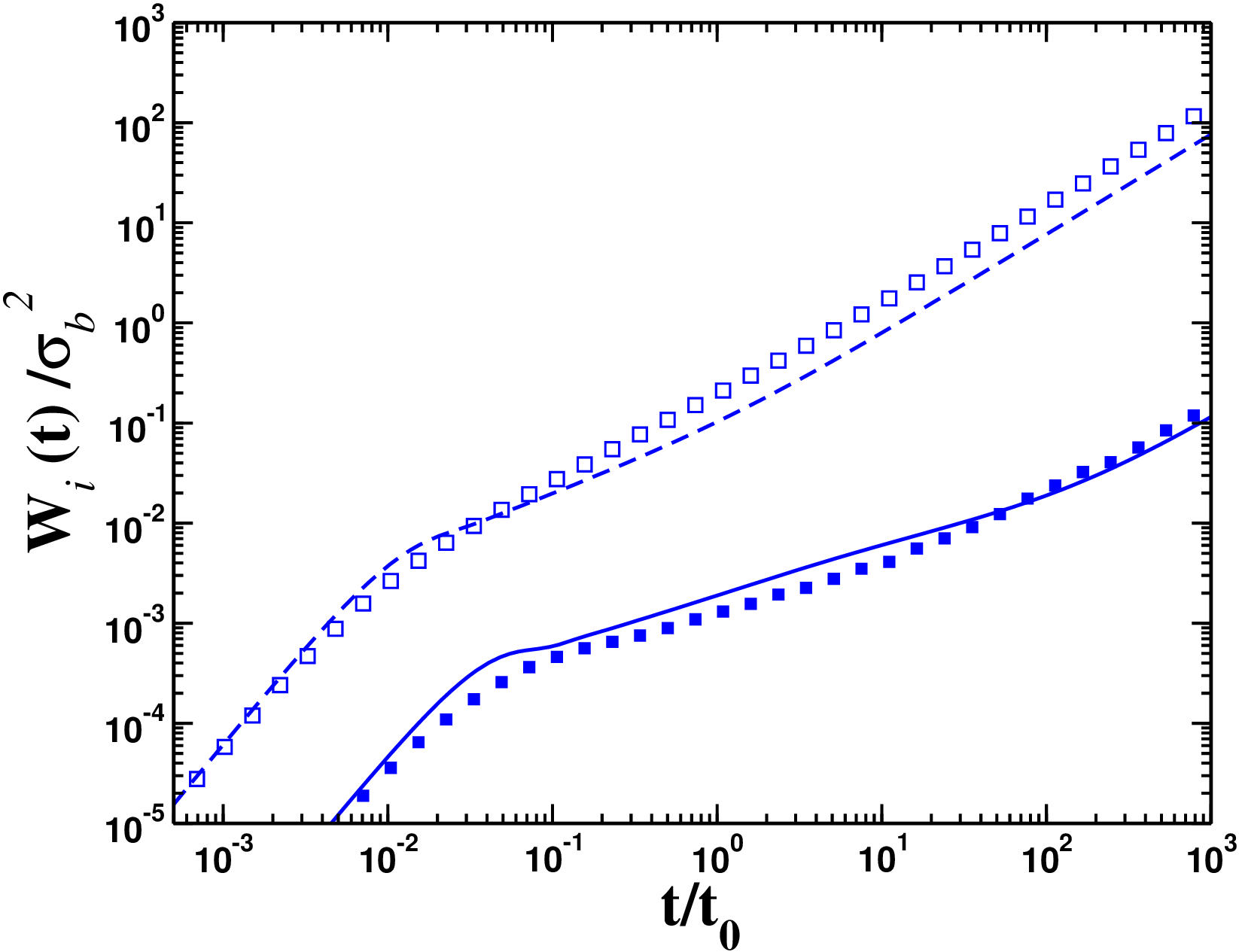}}
\subfigure[]{ \label{fig:3f}
\includegraphics[scale=.149]{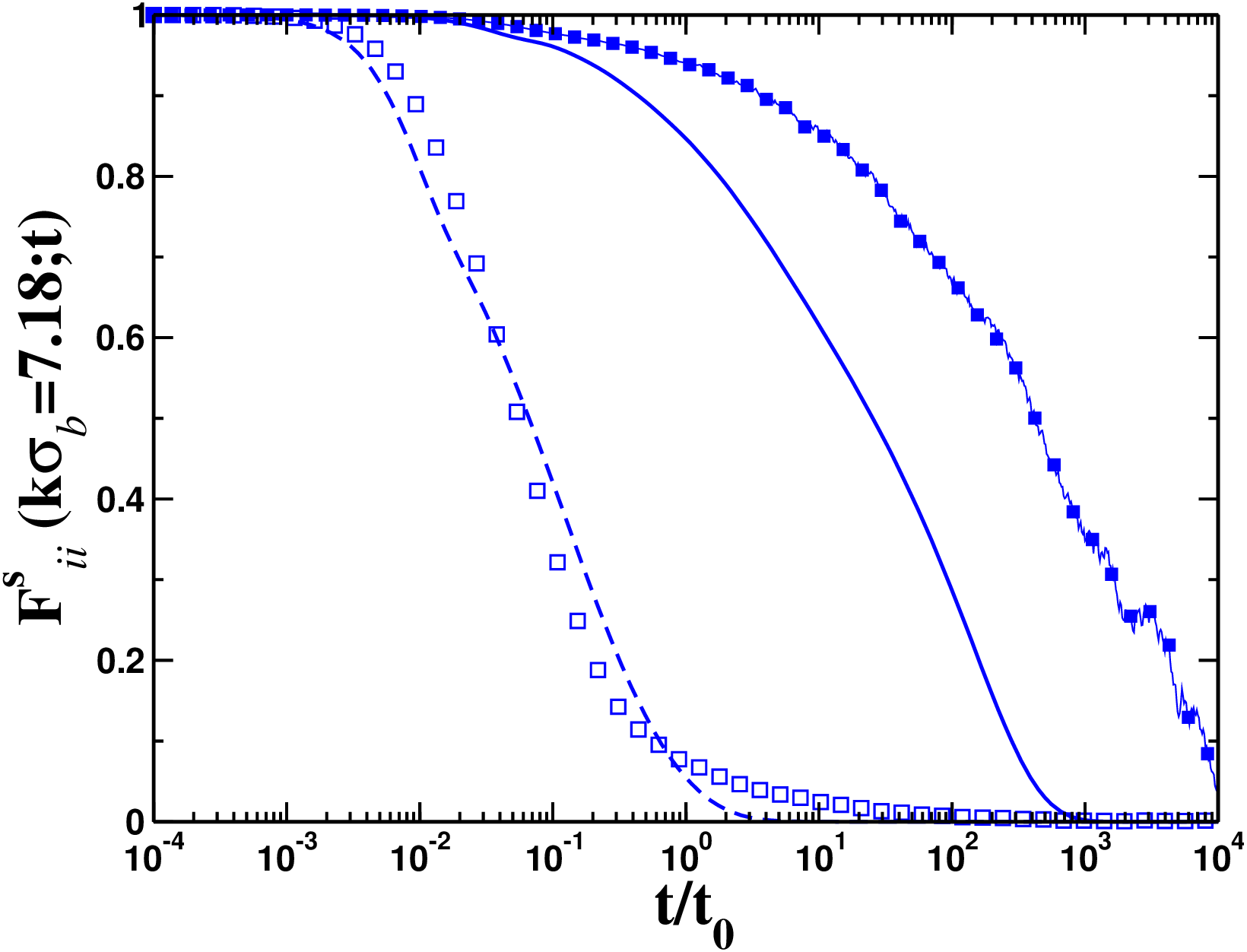}}
\caption{Theoretical (solid and dotted lines) and simulation (solid and open symbols)
results for the MSDs, $W_{i}(t)$, and ISFs $F^s_{i} (k=7.18;t)$ of
a highly asymmetric ($\delta=0.2$) binary mixture of HS,
for the three volume fractions, in the control parameter space $(\phi_b,\phi_s)$.
(a) and (b) Corresponds to the point \textbf{I}$=(0.45,0.05)$.
(c) and (d) Corresponds to \textbf{II}$=(0.45,0.2)$.
(e) and (f) Corresponds to \textbf{III}$=(0.60,0.05)$.
In all cases, solid symbols ($\CIRCLE$,\textcolor{red}{$\blacktriangle$},
\textcolor{blue}{$\blacksquare$}) and solid lines corresponds to the dynamical properties
of big particles, whereas empty symbols ($\Circle$, \textcolor{red}{$\bigtriangleup$},
\textcolor{blue}{$\square$})
and dashed lines accounts for small particles.}
\label{fig3}
\end{figure}

To test the accuracy of the resulting approximate theory, we have 
solved Eqs. (\ref{modwdtintdifec})-(\ref{fsmix}) with Eqs. (\ref{dkinetictheoryBIG}) 
and (\ref{dsmall}) for the MSDs, $W_i(t)$, and self-ISFs, $F^S_i(t)$, of the highly 
asymmetric binary HS mixture  ($\delta=0.2$) for three aforementioned state points in the 
high-concentration region of the control parameter space $(\phi_b,\phi_s)$, namely, 
\textbf{I}$=(0.45,0.05)$, \textbf{II}$=(0.45,0.2)$, and \textbf{III}$=(0.60,0.05)$. The 
corresponding predictions are compared in Fig. \ref{fig3} with their simulation counterparts. 
The first bird-eye conclusion of this comparison is that the theory provides quite a reasonable
representation of the simulation results, considering its approximate nature and the simplicity 
of its underlying approximations and simplifications. 

More quantitatively, we notice that the theoretical results exhibit the correct short- and
long-time behavior of $W_i(t)$, including the disparate time-scales of the dynamics of 
$W_b(t)$ and $W_s(t)$ when plotted as a function of the scaled time $t_m$ (defined in units
of $\sigma_b$ and $M_b$). This time-scale difference originates in the huge mass ratio
$M_s/M_b=\delta ^3 = 0.008$. This can be seen through the short-time (ballistic) solution of
Eq. (\ref{modwdtintdifec}), $<(\Delta \mathbf{r}_i (t))^2>  \approx [3 k_BT/M_i ]t^2$ which, 
expressed in terms of $W_{i}(t)=<(\Delta \mathbf{r}_i (t))^2>/6$, reads 
\begin{equation}
\dfrac{W_i(t_m)}{\sigma_b^2 } =\dfrac{1}{2}\left[ \dfrac{M_b}{M_{i}}\right] (t_m)^2, \label{wdt0shortmixscaled}
\end{equation}
where $t_m\equiv t/t^0_b$ and $t^0_b=\sigma_b\sqrt{M_b/k_BT}$. The ratio $[M_b/M_{i}]$ is
unity for the big particles ($i=b$) and is 0.008 for the small particles ($i=s$), thus 
explaining the aforementioned time-scale difference.

Figs. \ref{fig:3a}-\ref{fig:3b}, \ref{fig:3c}-\ref{fig:3d}
and \ref{fig:3e}-\ref{fig:3f} correspond, respectively, to the points
\textbf{I}, \textbf{II} and \textbf{III}. These figures illustrate the
effect on the dynamics of the system upon variations in the two control
parameters $\phi_b$ and $\phi_s$.
For instance, regarding the dynamics
of the system at the state point \textbf{I}$=(\phi_b=0.45,\phi_s=0.05)$, one 
notices that, besides the aforementioned time-scale difference in the MSDs of
both species observed in Fig. \ref{fig:3a}, the SCGLE results for   the ISFs 
of each species in  Fig. \ref{fig:3b} (solid and dashed lines) also reveals a
noticeable difference between the characteristic decay times of the ISFs of each 
species, although the one-step relaxation
pattern of each correlator is rather similar. The differences in the decay 
times are described by the $\alpha$-relaxation times, $\tau^{(\alpha)}_i$ ($i=b,s$),
defined here as $F^S_{i}(k\sigma_b=7.18;\tau^{(\alpha)}_i)=1/e$. These
features are consistent with the results obtained from MD simulations (solid
and open symbols).

Upon increasing $\phi_s$ from 0.05 to 0.2 we go from state point \textbf{I} to 
state point \textbf{II}$=(0.45,0.2)$, and  in Figs. \ref{fig:3c} and \ref{fig:3d}
we observe that the mobility of both species decrease. In addition, the crossover
from the ballistic to the diffusive regime occurs at shorter times compared with 
the situation
illustrated at \textbf{I}, and the
difference between the corresponding $\alpha$-relaxation times become larger.
The simulation results nicely confirm these trends.

A more interesting behavior is observed when we move from state point \textbf{I} to 
state point  \textbf{III}$=(0.6,0.05)$, this time by fixing $\phi_s=0.05$ and increasing
$\phi_b$ from 0.45 to 0.6, above the glass transition threshold $\phi^{(g)}\approx0.582$
of the monodisperse HS fluid. As illustrated in Figs. \ref{fig:3e}-\ref{fig:3f}, in addition
to the different time-scales of $W_b(t)$ and $W_s(t)$ derived from the disparate mass difference,
a further enhancement of this difference is now quite visible at long times, suggesting a strong 
disparity in the structural relaxation of the two species. For instance, a large disparity in the
mobility of each species, measured by the ratio $W_b(t_m=10^3)/W_s(t_m=10^3)\sim10^{-3}$, is observed.
Notice also the emergence of an incipient plateau in the MSD of the large particles, which is absent 
in the MSD of the small ones. 

This long-time dynamic asymmetry is also observed in the ISFs of each species, which
displays different relaxation patterns, characterized by a faster relaxation mechanism
for the small particles and a far slower relaxation of the large particles. Except for 
quantitative details, mostly manifested in $F^S_{bb}(k,t)$, the comparison with the simulation
data demonstrates that these theoretical predictions capture the essential phenomenology of the 
simulation results. 

\section{Concluding remarks}\label{sectionV}

In this paper we have proposed and tested an approximate 
but quantitative theoretical approach for the description
of the dynamics of \emph{fully equilibrated} atomic liquid
mixtures. Such framework was built on the exact time-evolution 
equations for the long-time dynamics of an atomic liquid, 
previously developed in Refs. \cite{atomic1,atomic2,atomic3}, 
which were complemented by a set of well-defined approximations,
including a Gaussian-like approximation that incorporates the 
correct ballistic short-time limit. The predictive accuracy of
the resulting theoretical tool was confirmed with the assistance 
of pertinent molecular dynamics simulations. The  general conclusion 
drawn from these numerical tests indicate a remarkable degree of 
reliability of the present SCGLE theory. 
Although its quantitative accuracy could be improved in several
manners, this was not the primary interest of the present work. 
Here we focused, instead, in illustrating the systematic use of our
theory in two representative and concrete examples. The first 
involved a simple but polydisperse HS liquid and the second a 
genuine and highly-asymmetric binary mixture of hard spheres. 

In both cases we expect that the present approximate theory
will evolve into a useful theoretical tool to model the properties 
of experimentally-relevant atomic liquid mixtures, such as molten 
salts and metallic alloys. This unification of the physics of colloidal
and atomic liquids clearly creates an opportunity to systematically transfer
much of the knowledge generated in the field of colloids to the understanding 
of complex atomic liquid mixtures and vice versa.
For instance, both examples discussed here actually derive from two separate 
projects involving model colloidal HS liquids, whose properties could in practice
be more easily simulated using event-driven molecular dynamics, rather than the
more natural Brownian dynamics  simulations. In the first example, since hard-sphere
colloids are usually polydisperse, theoretically describing polydispersity with the 
SCGLE theory and then testing the results with MD simulations is now a natural and 
handy modeling protocol, as discussed in more detail elsewhere \cite{patygabriel}. 
The same protocol is being followed in modeling the dynamics of genuine HS colloidal
mixtures with large size asymmetry, and discussed in more detail in separate work  \cite{todoss}.

Finally, another particularly relevant opportunity is represented by the possibility 
of connecting the advances in our understanding of the formation of colloidal glasses
and gels with the technologically relevant need to understand the formation of amorphous
solids by the cooling of complex glass- and gel-forming atomic liquids. Although this 
more ambitious project requires the development of the non-equilibrium version of the 
present SCGLE theory of equilibrium atomic liquids, the present work paves the way for
such developments. 

\vskip1cm

\section{Acknowledgments}
This work was supported by the Consejo Nacional de
Ciencia y Tecnolog\'{\i}a (CONACYT, M\'{e}xico) through
grants Nos. 242364, 182132, 237425, 358254, and FC-2015-2/1155,
and by the Universidad
de Guanajuato (through the Convocatoria Institucional para el
Fortalecimiento de la Excelencia Acad\'emica 2015).
L.F.E.A. and R.C.P. acknowledge financial support from
Secretar\'ia de Educaci\'on P\'ublica (SEP, M\'exico) through
Postdoctoral fellowship, PRODEP. 
P.M.M. and M.M.N. acknowledge the support of Secretar\'ia de
Educaci\'on P\'ublica through Postdoctoral fellowship, PRODEP.
L.F.E.A also acknowledge financial support from the
German Academic Exchange Service (DAAD) through the DLR-DAAD
programme under grant No. 212. R. C. P. also acknowledges the
financial support provided by the Marcos Moshinsky fellowship
2013 - 214 and the Alexander von Humboldt Foundation during his
stay at the University of D\"usseldorf in summer 2016.
The authors acknowledge  Thomas Voigtmann for interesting discussions.

\end{document}